\documentclass[amsmath,amssymb,aps,prb,twocolumn,nolongbibliography]{revtex4-1}
\usepackage[utf8]{inputenc}
\usepackage{graphicx}
\usepackage{dcolumn}
\usepackage{bm}
\usepackage{dsfont}
\usepackage{physics}
\usepackage{xcolor}
\usepackage{soul} 
\sethlcolor{cyan}
\usepackage{float}
\usepackage{orcidlink}

\begin{document}

\title{Orbital Angular Momentum Textures and Currents in a Discrete Helix: Equilibrium and Linear Response}

\author{Danny Cordova\orcidlink{0009-0005-5675-9395}}
 \affiliation{Departamento de F\'isica, Colegio de Ciencias e Ingenier\'ia, Universidad San Francisco de Quito, Diego de Robles y Via Interoceanica, Quito 17901, Ecuador}

\author{Bertrand Berche\orcidlink{0000-0002-4254-807X}}
\affiliation{Laboratoire de Physique et Chimie Théoriques, CNRS - Université de Lorraine, UMR 7019,
Nancy, France}

\author{Ernesto Medina\orcidlink{0000-0002-1566-0170}}
\affiliation{Departamento de F\'isica, Colegio de Ciencias e Ingeniería, Universidad San Francisco de Quito, Diego de Robles y Via Interoceánica, Quito 17901, Ecuador}
\affiliation{School of Molecular Sciences, Arizona State University, 551E University Dr, Tempe, AZ 85281, USA}

\date{\today}

\begin{abstract}
Recently, nonequilibrium orbital angular momentum in low-dimensional systems has attracted renewed attention. Here we introduce a minimal three-orbital tight-binding model for a single helical chain and show that chirality alone generates a momentum-dependent orbital-angular-momentum texture through Slater--Koster hybridization in the local basis $(p_r,p_\phi,p_z)$, without requiring atomic spin--orbit coupling. In the single-helix geometry, the radial orbital texture vanishes identically, while the azimuthal and longitudinal components remain finite and arise from the odd-in-momentum $(p_z,p_r)$ and $(p_r,p_\phi)$ sectors. As a result, the equilibrium average orbital texture vanishes by parity, although persistent-like orbital angular momentum currents may still exist and imply chirality-dependent end magnetization in a finite helix. Under an applied longitudinal electric field, the system develops a finite orbital Edelstein response, whereas the projected longitudinal orbital-current conductivity vanishes in the linear regime by parity. When spin degrees of freedom are included, the orbital texture acts as a source of spin polarization through orbital-to-spin transduction. The resulting spin response is controlled by orbital overlap scales much larger than the bare relativistic spin--orbit scale, making it a stronger candidate for spin injection than the conventional spin Edelstein mechanism. These results identify chirality as the minimal microscopic ingredient for generating orbital angular momentum response in one-dimensional systems and support an orbital route to spin selectivity in chiral conductors.

\end{abstract}

  \maketitle

\section{Introduction}

The generation and control of angular momentum in electronic transport have become central themes in spintronics and, more broadly, in the physics of low-dimensional systems. While early developments focused primarily on spin degrees of freedom within spintronics\cite{vzutic2004spintronics}, recent advances in orbitronics have demonstrated that orbital angular momentum can constitute an independent and efficient carrier of information \cite{Cysne2025Orbitronics,Wang2024OrbitronicsReview}. Orbital currents can be generated even in materials with weak spin--orbit interaction and even with no atomic SOC at all and subsequently converted into spin accumulation or magnetization through interfacial coupling, offering new routes toward low-dissipation devices \cite{Go2020OrbitalTorque,Sala2022OrbitalConversion}. In particular, orbital textures emerging from crystal symmetry, hybridization between atomic orbitals, and Berry-phase structure in momentum space have been shown to produce current-induced responses such as the orbital Hall and orbital Edelstein effects, which may precede or even dominate their spin counterparts \cite{Yoda2018OrbitalEdelstein,Busch2023OrbitalHall}.

Chirally induced spin selectivity (CISS) represents one of the most striking manifestations of angular momentum selectivity in transport.  Experiments across molecular junctions, organic assemblies, and chiral crystalline materials consistently report substantial spin polarization generated in the absence of magnetic materials or strong magnetic fields \cite{bloom2024chiral,Gupta2026CISSReview}. These observations are supported by a curated body of literature detailing spin filtering in self-assembled monolayers \cite{ray1999asymmetric,gohler2011spin}, DNA \cite{guo2012spin,naaman2022chiral}, and chiral crystalline frameworks \cite{inui2020chirality,calavalle2022gate,kettner2018chirality}. Device architectures and spectroscopic measurements have further highlighted robust magnetoresistance and gate-tunable polarization \cite{aragones2022magnetoresistive,liu2020linear}. 

Most theoretical descriptions attribute this behavior to the combined action of spin--orbit interaction and structural chirality, whereby the lack of inversion symmetry enables spin-dependent scattering \cite{aiello2022chirality,naaman2019chiral}. However, several conceptual difficulties remain. The intrinsic spin--orbit coupling in many experimentally relevant systems is comparatively weak \cite{varela2016effective}, and reciprocity relations impose strong constraints on spin polarization in coherent two-terminal transport under time-reversal symmetry \cite{nitzan2001electron,yang2020detecting}. As a consequence, tunneling processes, environmental decoherence, or other effective mechanisms that break time-reversal symmetry are frequently required to reproduce experimentally observed polarization magnitudes \cite{mena2024Minimal,diaz2018thermal,michaeli2019origin,Bedoya2026}. These considerations have led to an ongoing debate regarding the microscopic origin of CISS and the relative importance of electronic, vibrational, and environmental effects \cite{Miwa2024VibrationalCISS,fransson2020vibrational}.

Orbital degrees of freedom have also been considered in several approaches to CISS, where helical hybridization, decoherence, and correlation effects were shown to generate angular momentum selectivity beyond purely spin--orbit driven mechanisms \cite{Gutierrez2012HelicalTB,matityahu2016spin,fransson2020vibrational}. This theoretical foundation relies heavily on mapping spin transport onto explicit helical potentials \cite{yeganeh2009chiral}, evaluating symmetry-derived tight-binding implementations \cite{geyer2019chirality,PMendieta}, and incorporating rigorous treatments of non-unitary scattering and dephasing \cite{Cattena2010}.

An alternative viewpoint is that chirality may primarily generate orbital angular momentum rather than spin polarization itself, a concept linked to observations of electron-optic dichroism \cite{farago1980spin,mayer1995experimental,varela2024optical}. Recent theoretical work has demonstrated that electric currents flowing through chiral crystals produce orbital magnetization responses whose sign is controlled by structural handedness, even in the absence of atomic spin--orbit coupling \cite{Yoda2015HelicalMagnetization}. This phenomenon, closely related to the orbital Edelstein effect, establishes that geometry alone can generate nonequilibrium angular momentum textures under bias \cite{Yoda2018OrbitalEdelstein}. From this perspective, spin polarization arises as a secondary process resulting from the conversion of orbital to spin angular momentum.

In this work, we introduce a tight-binding model in which orbital angular momentum texture and currents arise directly from the helical geometry of a single chain and the application of a voltage bias.  The mechanism originates from Slater--Koster hybridization \cite{harrison1989electronic} between longitudinal and transverse $p$ orbitals enforced by screw symmetry. No atomic spin--orbit interaction is required for the generation of orbital transport. Instead, a finite bias produces a nonequilibrium orbital texture through a linear-response mechanism analogous to an orbital Edelstein response. Within this framework, chirality acquires a precise microscopic role. The helical structure prevents inversion-related cancellation of inter-orbital overlaps and generates an odd and even-in-momentum hybridization term in the Bloch Hamiltonian \cite{kochan2017model}. The resulting Bloch states carry a finite atomic orbital angular momentum, whose sign is determined by the structure's handedness.

Spin accumulation nevertheless requires conversion of orbital angular momentum into spin angular momentum. 
 We discuss possible orbital-to-spin conversion pathways in the bulk of the molecule, introducing the Spin-Orbit-Coupling, which can reconcile the orbital mechanism proposed here with the experimentally observed spin polarization. 
 
 The paper is structured as follows: We first introduce a three-orbital per-site model. The helical structure mixes the orbitals, producing an effective on-site orbital-angular-momentum texture that can be computed analytically. We then compute the orbital texture and currents. Following, we compute the linear response to bias resulting in the orbital Edelstein response $\chi_L$ and the orbital current conductivity $\sigma_L$. The bulk coupling of the orbital current to the spin is computed by introducing the SOC\cite{varela2016effective}. This transduction is compared with the bare spin-Edelstein current, showing that the orbital-to-spin transduction is more effective, even in the presence of meV SOC. We conclude with a discussion of order-of-magnitude estimates and other possible mechanisms for orbital-to-spin conversion.

 \begin{figure} [ht]
    \centering
    \includegraphics[width=0.95\linewidth]{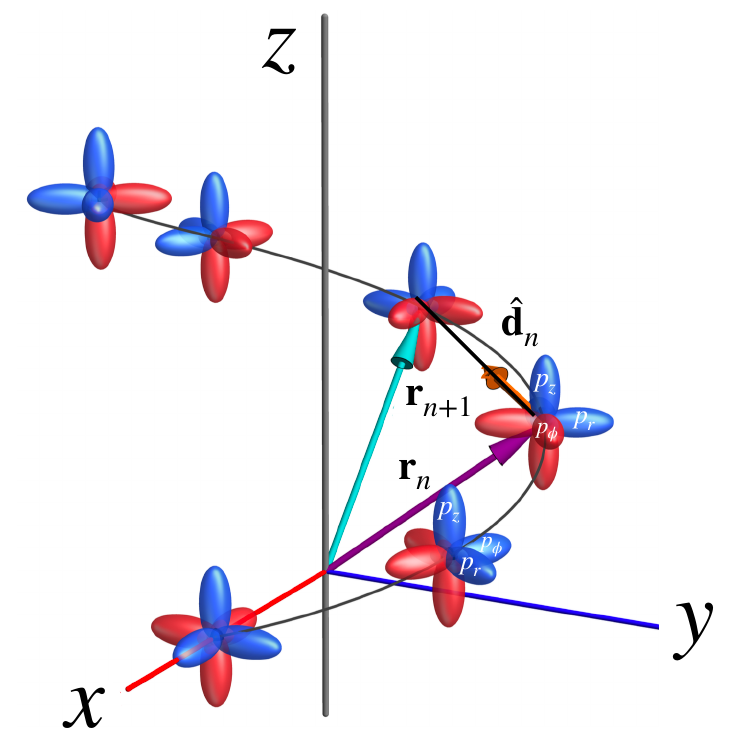}
    \caption{(Color online). Model of the helix with the nearest neighbor sites ${\bf r}_n$ and ${\bf r}_{n+1}$. The included orbitals per site are $p_r,p_{\phi},p_z$ as depicted. The vector $\hat{\bf d}_n$ joins two adjacent sites. The orientation chosen for the orbital corresponds to a single strand DNA like structure.}
    \label{fig:helixmodel}
\end{figure}
    \label{fig:unit_projections}
 
\section{The Model}


We introduce a minimal three-orbital per-site tight-binding model for a single helical  (see Fig.~\ref{fig:helixmodel}). In contrast with the reduced two-orbital description, this formulation retains the local orbital set explicitly
\[
\left\{ p_r(n),\, p_\phi(n),\, p_z(n) \right\},
\]
for the local $p_x=p_r$, $p_y=p_{\phi}$ and $p_z=p_z$, and therefore allows for a consistent description of the full local orbital-angular-momentum texture of the Bloch states. The model Hamiltonian is then
\begin{equation}
H =
\sum_n \sum_{\alpha=r,\phi,z}
\epsilon_\alpha\, c_{n\alpha}^\dagger c_{n\alpha}
+
\sum_n \left( H_{n,n+1} + H_{n+1,n} \right),
\label{eq:H_real_space_full}
\end{equation}
with nearest-neighbor hopping term
\begin{equation}
H_{n,n+1}
=
\sum_{\alpha,\beta\in\{r,\phi,z\}}
t_{\alpha\beta}\,
c_{n+1,\alpha}^\dagger c_{n,\beta}.
\label{eq:H_nn_full}
\end{equation}

\label{sec:model}
We consider a single helical chain with sites at
\begin{equation}
\mathbf r_n=
\bigl(
R\cos(n\varphi),\,
R\sin(n\varphi),\,
nh
\bigr),
\end{equation}
where \(R\) is the helix radius, \(\varphi\) is the azimuthal step between neighboring sites, and \(h\) is the axial rise per site. The nearest-neighbor distance is
\begin{equation}
a=\left|\mathbf r_{n+1}-\mathbf r_n\right|
=
\sqrt{4R^2\sin^2\!\left(\frac{\varphi}{2}\right)+h^2}.
\end{equation}
At each site, we use the local cylindrical basis
\begin{eqnarray}
\hat{\mathbf r}_n=
(\cos n\varphi,\sin n\varphi,0),&\quad&
\hat{\boldsymbol\phi}_n=
(-\sin n\varphi,\cos n\varphi,0),\nonumber\\
\hat{\mathbf z}&=&(0,0,1),
\end{eqnarray}
and the corresponding local orbitals
\((p_r,p_\phi,p_z)\).
The bond vector from site \(n\) to site \(n+1\) is
\begin{align}
\mathbf d_n
&=\mathbf r_{n+1}-\mathbf r_n \nonumber\\
&=
R(\cos\varphi-1)\,\hat{\mathbf r}_n
+
R\sin\varphi\,\hat{\boldsymbol\phi}_n
+
h\,\hat{\mathbf z}.
\end{align}
Its unit vector can be written as
\begin{equation}
\hat{\mathbf d}_n
=
\alpha\,\hat{\mathbf r}_n
+
\beta\,\hat{\boldsymbol\phi}_n
+
\gamma\,\hat{\mathbf z},
\end{equation}
with
\begin{eqnarray}
\alpha&=&\frac{R(\cos\varphi-1)}{a},
\qquad
\beta=\frac{R\sin\varphi}{a},
\qquad\nonumber\\
\gamma&=&\frac{h}{a},
\qquad
\alpha^2+\beta^2+\gamma^2=1.
\end{eqnarray}
It is useful to note explicitly that
\begin{equation}
\alpha=-\frac{2R}{a}\sin^2\!\left(\frac{\varphi}{2}\right),
\quad
\beta=\frac{2R}{a}\sin\!\left(\frac{\varphi}{2}\right)\cos\!\left(\frac{\varphi}{2}\right),
\quad
\gamma=\frac{h}{a},
\end{equation}
so the sign carried by \(\alpha\) and the sign carried by \(\beta\) remain visible.

The local frame at site \(n+1\) is obtained from that at site \(n\) by a rotation through \(\varphi\) around \(\hat{\mathbf z}\):
\begin{eqnarray}
\hat{\mathbf r}_{n+1}
&=&
\cos\varphi\,\hat{\mathbf r}_n
+
\sin\varphi\,\hat{\boldsymbol\phi}_n,
\nonumber\\
\hat{\boldsymbol\phi}_{n+1}
&=&
-\sin\varphi\,\hat{\mathbf r}_n
+
\cos\varphi\,\hat{\boldsymbol\phi}_n.
\end{eqnarray}
Therefore,
\begin{equation}
\hat{\mathbf r}_{n+1}\cdot\hat{\mathbf d}_n=-\alpha,
\qquad
\hat{\boldsymbol\phi}_{n+1}\cdot\hat{\mathbf d}_n=\beta,
\qquad
\hat{\mathbf z}\cdot\hat{\mathbf d}_n=\gamma,
\end{equation}
and the needed axis overlaps are
\begin{align}
\hat{\mathbf r}_n\cdot\hat{\mathbf r}_{n+1}&=\cos\varphi,
&
\hat{\mathbf r}_n\cdot\hat{\boldsymbol\phi}_{n+1}&=-\sin\varphi,
\nonumber\\
\hat{\boldsymbol\phi}_n\cdot\hat{\mathbf r}_{n+1}&=\sin\varphi,
&
\hat{\boldsymbol\phi}_n\cdot\hat{\boldsymbol\phi}_{n+1}&=\cos\varphi.
\end{align}

For nearest neighbors, the Slater--Koster matrix element between local \(p\)-orbitals directed along unit vectors \(\hat{\mathbf e}_\mu(n)\) and \(\hat{\mathbf e}_\nu(n+1)\) is
\begin{equation}
t_{\mu\nu}
=
V_{pp\pi}\,
\hat{\mathbf e}_\mu(n)\cdot \hat{\mathbf e}_\nu(n+1)
+
\Delta\,
\bigl[\hat{\mathbf e}_\mu(n)\cdot\hat{\mathbf d}_n\bigr]
\bigl[\hat{\mathbf e}_\nu(n+1)\cdot\hat{\mathbf d}_n\bigr],
\end{equation}
where $\Delta \equiv V_{pp\sigma}-V_{pp\pi}$.

In the basis \((p_r,p_\phi,p_z)\), the forward nearest-neighbor hopping matrix \(T\) from site \(n\) to site \(n+1\) is
\begin{equation}
T=
\begin{pmatrix}
t_{rr} & t_{r\phi} & t_{rz}\\
t_{\phi r} & t_{\phi\phi} & t_{\phi z}\\
t_{zr} & t_{z\phi} & t_{zz}
\end{pmatrix},
\end{equation}
with
\begin{align}
\label{sk_parameters}
t_{rr}&=V_{pp\pi}\cos\varphi-\Delta\,\alpha^2,\nonumber\\
t_{r\phi}&=-V_{pp\pi}\sin\varphi+\Delta\,\alpha\beta,\nonumber\\
t_{rz}&=\Delta\,\alpha\gamma,\nonumber\\
t_{\phi r}&=V_{pp\pi}\sin\varphi-\Delta\,\alpha\beta=-t_{r\phi},\nonumber\\
t_{\phi\phi}&=V_{pp\pi}\cos\varphi+\Delta\,\beta^2,\nonumber\\
t_{\phi z}&=\Delta\,\beta\gamma,\nonumber\\
t_{zr}&=-\Delta\,\alpha\gamma=-t_{rz},\nonumber\\
t_{z\phi}&=\Delta\,\beta\gamma=t_{\phi z},\nonumber\\
t_{zz}&=V_{pp\pi}+\Delta\,\gamma^2.
\end{align}
Thus, the sign relations
$t_{zr}=-t_{rz},~t_{r\phi}=-t_{\phi r},~ t_{z\phi}=t_{\phi z}$
follow directly from the bond geometry and the right-handed cylindrical basis. Thus, the helical geometry naturally produces antisymmetric hopping in the $(p_z,p_r)$ and $(p_r,p_\phi)$ sectors, while the $(p_z,p_\phi)$ coupling remains symmetric. 

Although the full Slater--Koster construction in the local
\((p_r,p_\phi,p_z)\) basis contains several inter-orbital hopping channels,
not all of them encode {\it chirality} in the same way. In particular,
the \(r\!-\!\phi\) mixing is already present in a planar curved geometry and
mainly reflects the rotation of the local cylindrical frame from site to site.
By contrast, the \(r\!-\!z\) and \(\phi\!-\!z\) couplings require a finite pitch
and are therefore tied to the genuinely three-dimensional helical structure.

The final form of the real-space tight-binding Hamiltonian is then
\begin{equation}
H
=
\sum_n \mathbf c_n^\dagger E\,\mathbf c_n
+
\sum_n
\left(
\mathbf c_n^\dagger T\,\mathbf c_{n+1}
+
\mathbf c_{n+1}^\dagger T^\dagger \mathbf c_n
\right).
\end{equation}
where
\begin{equation}
\mathbf c_n=
\begin{pmatrix}
c_{n,r}\\
c_{n,\phi}\\
c_{n,z}
\end{pmatrix},
\qquad
E=
\begin{pmatrix}
\varepsilon_r & 0 & 0\\
0 & \varepsilon_\phi & 0\\
0 & 0 & \varepsilon_z
\end{pmatrix},
\end{equation}
where \(\varepsilon_r,\varepsilon_\phi,\varepsilon_z\) are the onsite energies in the local basis.

\subsection{Bloch Hamiltonian}

Using
\begin{equation}
\mathbf c_n=\frac{1}{\sqrt N}\sum_k e^{ikna}\,\mathbf c_k,
\end{equation}
the Hamiltonian becomes

\begin{equation}
H
=
\sum_k
{\bf c}_k^\dagger\, \mathcal H(k)\, {\bf c}_k,
\qquad
{\bf c}_k =
\begin{pmatrix}
c_{k,r} \\
c_{k,\phi} \\
c_{k,z}
\end{pmatrix},
\label{eq:H_bloch_full}
\end{equation}
with
\begin{equation}
H(k)={E}+Te^{ika}+T^\dagger e^{-ika}.
\end{equation}

Since \(T\) is real but not symmetric,
\begin{equation}
H(k)=E+(T+T^T)\cos(ka)+i(T-T^T)\sin(ka),
\end{equation}
which makes the even and odd parts under \(k\to -k\) explicit:
\begin{eqnarray}
H_{\rm even}(k)&=&E+(T+T^T)\cos(ka),
\nonumber\\
H_{\rm odd}(k)&=&i(T-T^T)\sin(ka).
\end{eqnarray}
Therefore, the full three-orbital Hamiltonian naturally contains both even and odd orbital hybridization channels. This will allow the Bloch states to develop a full local orbital-angular-momentum vector texture, with different components controlled by different inter-orbital coherences.

In components,
\begin{equation}
H(k)=
\begin{pmatrix}
A_k & iX_k & iY_k\\
-iX_k & B_k & Z_k\\
-iY_k & Z_k & C_k
\end{pmatrix},
\label{eq:Hk-compact}
\end{equation}
where
\begin{align}
A_k&=\varepsilon_r+2t_{rr}\cos(ka),\\
B_k&=\varepsilon_\phi+2t_{\phi\phi}\cos(ka),\\
C_k&=\varepsilon_z+2t_{zz}\cos(ka),\\
X_k&=2t_{r\phi}\sin(ka),\\
Y_k&=2t_{rz}\sin(ka),\\
Z_k&=2t_{\phi z}\cos(ka).
\end{align}
Equivalently,
\begin{eqnarray}
H_{\rm even}(k)&=&
\begin{pmatrix}
A_k & 0 & 0\\
0 & B_k & Z_k\\
0 & Z_k & C_k
\end{pmatrix},
\nonumber\\
H_{\rm odd}(k)&=&
\begin{pmatrix}
0 & iX_k & iY_k\\
-iX_k & 0 & 0\\
-iY_k & 0 & 0
\end{pmatrix}.
\end{eqnarray}
The band energies are obtained from
$\det\!\bigl[H(k)-\lambda I\bigr]=0$.
For the matrix in Eq.~\eqref{eq:Hk-compact}, the characteristic polynomial is
\begin{equation}
\lambda^3-c_1(k)\lambda^2+c_2(k)\lambda-c_3(k)=0,
\end{equation}
with
\begin{align}
c_1(k)&=A_k+B_k+C_k,\nonumber\\
c_2(k)&=A_kB_k+A_kC_k+B_kC_k-X_k^2-Y_k^2-Z_k^2,\nonumber\\
c_3(k)&=A_kB_kC_k+2X_kY_kZ_k-A_kZ_k^2-B_kY_k^2-C_kX_k^2.
\end{align}
Solving the cubic equation with the definitions
\begin{equation}
p_k=c_2(k)-\frac{c_1(k)^2}{3},
\qquad
q_k=-\frac{2c_1(k)^3}{27}+\frac{c_1(k)c_2(k)}{3}-c_3(k),
\end{equation}
the three exact band energies may be written as
\begin{align}
    E_\nu(k) &= \frac{c_1(k)}{3} + 2\sqrt{-\frac{p_k}{3}}\, \cos\bigg[ \frac{1}{3} \arccos\!\left( \frac{3q_k}{2p_k}\sqrt{-\frac{3}{p_k}} \right) \nonumber \\
    &\quad -\frac{2\pi (\nu-1)}{3} \bigg],
\end{align}
with $\nu=1,2,3$, whenever the three roots are distinct. Since \(H(k)\) is manifestly Hermitian, all three energies are real. 

\subsection{Exact Bloch eigenfunctions}

It is convenient to remove the explicit factors of \(i\) from the first row and column by the unitary transformation
\begin{equation}
\widetilde H(k)=U^\dagger H(k)U
=
\begin{pmatrix}
A_k & X_k & Y_k\\
X_k & B_k & Z_k\\
Y_k & Z_k & C_k
\end{pmatrix}.
\end{equation}
with $U=\mathrm{diag}(i,1,1)$.
This matrix is real and symmetric. For a given eigenvalue \(E_\nu(k)\), one convenient unnormalized eigenvector of \(\widetilde H(k)\) is
\begin{equation}
\widetilde v_{\nu}(k)\propto
\begin{pmatrix}
(B_k-E_\nu)(C_k-E_\nu)-Z_k^2\\
Y_k Z_k - X_k(C_k-E_\nu)\\
X_k Z_k - Y_k(B_k-E_\nu)
\end{pmatrix}.
\end{equation}
Transforming back to the original local basis gives the Bloch spinor
\begin{equation}
u_{\nu}(k)=U\,\widetilde v_{\nu}(k)
=
\frac{1}{\mathcal N_{\nu}(k)}
\begin{pmatrix}
i\Big[(B_k-E_\nu)(C_k-E_\nu)-Z_k^2\Big]\\
Y_k Z_k - X_k(C_k-E_\nu)\\
X_k Z_k - Y_k(B_k-E_\nu)
\end{pmatrix},
\label{eq-generalBlochSpinor}
\end{equation}
with normalization
\begin{eqnarray}
\mathcal N_{\nu}(k)^2&=&
\Big[(B_k-E_\nu)(C_k-E_\nu)-Z_k^2\Big]^2\nonumber\\
&+&\Big[Y_k Z_k - X_k(C_k-E_\nu)\Big]^2\nonumber\\
&+&\Big[X_k Z_k - Y_k(B_k-E_\nu)\Big]^2.
\end{eqnarray}
Note that $A_k,B_k,C_k,X_k, Y_k, Z_k$ are real-valued, so of the three entries in the Bloch functions, only the first entry is imaginary.

The corresponding Bloch eigenfunction is
\begin{widetext}
\begin{equation}
|\psi_{{\nu}k}\rangle
=
\frac{1}{\sqrt N}
\sum_n e^{ikna}
\left[
u_{{\nu},r}(k)\,|n,p_r\rangle
+
u_{{\nu},\phi}(k)\,|n,p_\phi\rangle
+
u_{{\nu},z}(k)\,|n,p_z\rangle
\right].
\end{equation}
\end{widetext}
At the time-reversal-invariant momenta $k=0$ and $k=\pi/a$, one has $\sin(ka)=0$, hence $X_k=Y_k=0$, and the $p_r$ orbital decouples from the $(p_\phi,p_z)$ sector. In that case the Hamiltonian reduces to a $1\oplus 2$ block structure, with one purely radial band and two bands obtained from the even-in-$k$ mixing in the $(p_\phi,p_z)$ sector.

\begin{figure}[ht]
    \centering
    \includegraphics[width=1\linewidth]{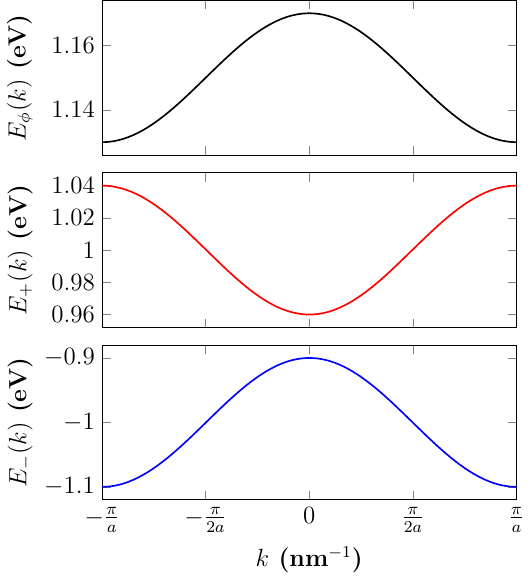}
    \caption{ Band structure of the reduced $t_{rz}$-only model. The azimuthal band $E_\phi(k)$ (online black) remains fully decoupled, while the $p_r-p_z$ sector hybridizes through the odd-in-momentum coupling $Y_k$, yielding the split branches $E_+(k)$ (online red) and $E_-(k)$ (online blue). The parameters used are reported in the appendix.}
    \label{fig:orbital_decoupling}
\end{figure}

\subsection{Illustrative reduced limit: only \(t_{zr}=-t_{rz}\) kept among inter-orbital hoppings}
\label{subsec:reduced-limit}

For illustration, it is useful to consider a reduced model in which the only nonzero inter-orbital hopping is the \(p_r\)--\(p_z\) one,
\begin{equation}
t_{r\phi}=t_{\phi r}=0,
\qquad
t_{\phi z}=t_{z\phi}=0,
\qquad
t_{zr}=-t_{rz}\neq 0.
\end{equation}
This is not the generic Slater--Koster helix, but it isolates the sector that later controls the odd-in-\(k\) orbital texture and provides a simple closed-form limit.

In this case, the hopping matrix becomes
\begin{equation}
T_{\rm red}=
\begin{pmatrix}
t_{rr} & 0 & t_{rz}\\
0 & t_{\phi\phi} & 0\\
-t_{rz} & 0 & t_{zz}
\end{pmatrix},
\end{equation}
and the Bloch Hamiltonian is
\begin{equation}
H_{\rm red}(k)=
\begin{pmatrix}
A_k & 0 & iY_k\\
0 & B_k & 0\\
-iY_k & 0 & C_k
\end{pmatrix},
\label{eq:Hk-reduced}
\end{equation}
with
\begin{eqnarray}
A_k&=&\varepsilon_r+2t_{rr}\cos(ka),
\nonumber\\
B_k&=&\varepsilon_\phi+2t_{\phi\phi}\cos(ka),
\nonumber\\
C_k&=&\varepsilon_z+2t_{zz}\cos(ka),
\nonumber\\
Y_k&=&2t_{rz}\sin(ka).
\end{eqnarray}
The \(p_\phi\) orbital decouples completely, so one band is
\begin{equation}
E_\phi(k)=B_k,
\qquad
u_\phi(k)=
\begin{pmatrix}
0\\
1\\
0
\end{pmatrix}.
\label{eq-decoupling}
\end{equation}
The remaining two bands come from the \(p_r\)--\(p_z\) block,
\begin{equation}
\begin{pmatrix}
A_k & iY_k\\
-iY_k & C_k
\end{pmatrix},
\end{equation}
and their energies are
\begin{equation}
E_\pm(k)
=
\frac{A_k+C_k}{2}
\pm
\sqrt{
\left(\frac{A_k-C_k}{2}\right)^2+Y_k^2
}.
\label{eq:Epm-reduced}
\end{equation}
The bands are plotted in Fig.~\ref{fig:orbital_decoupling}.
To write the eigenvectors, define the mixing angle \(\theta_k\) by
\begin{eqnarray}
\cos(2\theta_k)&=&
\frac{C_k-A_k}{\sqrt{(C_k-A_k)^2+4Y_k^2}}\nonumber,\\
\sin(2\theta_k)&=&
\frac{2Y_k}{\sqrt{(C_k-A_k)^2+4Y_k^2}}.
\label{eq:theta-def}
\end{eqnarray}
Then a convenient choice of normalized Bloch spinors is
\begin{eqnarray}
u_+(k)&=&
\begin{pmatrix}
i\cos\theta_k\\
0\\
\sin\theta_k
\end{pmatrix},\nonumber\\
u_-(k)&=&
\begin{pmatrix}
-i\sin\theta_k\\
0\\
\cos\theta_k
\end{pmatrix},\nonumber\\
u_\phi(k)&=&
\begin{pmatrix}
0\\
1\\
0
\end{pmatrix},
\end{eqnarray}
corresponding to the three energies \(E_+(k)\), \(E_-(k)\), and \(E_\phi(k)\), respectively.

The associated Bloch eigenfunctions are
\begin{equation}
|\psi_{\pm,k}\rangle
=
\frac{1}{\sqrt N}
\sum_n e^{ikna}
\left[
u_{\pm,r}(k)\,|n,p_r\rangle
+
u_{\pm,z}(k)\,|n,p_z\rangle
\right],
\end{equation}
and
\begin{equation}
|\psi_{\phi,k}\rangle
=
\frac{1}{\sqrt N}
\sum_n e^{ikna}
|n,p_\phi\rangle.
\end{equation}
This reduced limit makes two points: First, the \(p_r\)--\(p_z\) mixing is entirely odd in momentum because it is proportional to $
Y_k=2t_{rz}\sin(ka)$.
Second, at the time-reversal invariant momenta \(k=0\) and \(k=\pi/a\), where \(\sin(ka)=0\), the mixing vanishes, and the eigenstates become pure local orbitals:
\begin{equation}
u_+(k),u_-(k)\ \to\ \text{pure }p_r\text{ or }p_z,
\qquad
u_\phi(k)=
\begin{pmatrix}
0\\1\\0
\end{pmatrix}.
\end{equation}
This is the simplest limit in which the odd-in-\(k\) structure generated by the helical geometry can be displayed in closed form.

\subsection{Orbital angular momentum texture}

In the full three-orbital basis $\left\{|p_r\rangle,\,
|p_\phi\rangle,\,
|p_z\rangle
\right\}$ the local orbital angular momentum is a vector quantity. Its components arise from inter-orbital coherences among the three $p$ orbitals. In the local cylindrical basis, the orbital angular momentum operators are
\begin{align}
\hat L_r
&=
i\hbar
\left(
|p_\phi\rangle\langle p_z|
-
|p_z\rangle\langle p_\phi|
\right),
\label{eq:Lr_operator}
\\
\hat L_\phi
&=
i\hbar
\left(
|p_z\rangle\langle p_r|
-
|p_r\rangle\langle p_z|
\right),
\label{eq:Lphi_operator_full}
\\
\hat L_z
&=
i\hbar
\left(
|p_r\rangle\langle p_\phi|
-
|p_\phi\rangle\langle p_r|
\right).
\label{eq:Lz_operator}
\end{align}
These operators act locally on each site. Although the microscopic Hamiltonian contains only inter-site hopping terms, the Bloch eigenstates develop on-site inter-orbital coherences and, therefore, carry a finite local orbital-angular-momentum texture.

Let $|u_\nu(k)\rangle$ denote a normalized eigenstate of the Bloch Hamiltonian $\mathcal H(k)$:
\begin{equation}
\mathcal H(k)\, |u_\nu(k)\rangle = E_\nu(k)\, |u_\nu(k)\rangle,
\label{eq:eigenproblem_full}
\end{equation}
with band index $\nu=1,2,3$. In the local orbital basis, we write
\begin{equation}
|u_\nu(k)\rangle
=
a_r^{(\nu)}(k)\, |p_r\rangle
+
a_\phi^{(\nu)}(k)\, |p_\phi\rangle
+
a_z^{(\nu)}(k)\, |p_z\rangle,
\label{eq:eigenvector_components}
\end{equation}
where the values of $a^{(\nu)}_{\alpha}$ can be read off from Eq.~\ref{eq-generalBlochSpinor}. Also $\left| a_r^{(\nu)}(k) \right|^2
+
\left| a_\phi^{(\nu)}(k) \right|^2
+
\left| a_z^{(\nu)}(k) \right|^2
=1$.
The local orbital-angular-momentum texture carried by the Bloch state is then
\begin{equation}
\mathbf L_\nu(k)
=
\left(
\langle \hat L_r \rangle_\nu(k),\,
\langle \hat L_\phi \rangle_\nu(k),\,
\langle \hat L_z \rangle_\nu(k)
\right),
\label{eq:L_vector_texture}
\end{equation}
with components
\begin{align}
\langle \hat L_{\alpha} \rangle_\nu(k)
&=
\langle u_\nu(k) | \hat L_{\alpha} | u_\nu(k) \rangle
=
i\hbar
\varepsilon_{\alpha\beta\gamma} a_\beta^{(\nu)*}(k)\, a_{\gamma}^{(\nu)}(k)
 \label{eq:L_texture}
 \end{align}
So we have that
 \begin{eqnarray}
    \langle \hat{L}_{\phi} \rangle_{\nu}(k) &=& i\hbar [ 2i \mathrm{Im}( a^{(\nu)*}_{z}(k) a^{(\nu)}_{r}(k) ) ]\nonumber\\
    &=& -2\hbar \mathrm{Im} [ a^{(\nu)*}_{z}(k) a^{(\nu)}_{r}(k) ]
\end{eqnarray}
and
\begin{eqnarray}
    \langle \hat{L}_{z} \rangle_{\nu}(k) &=& i\hbar [ 2i \mathrm{Im}( a^{(\nu)*}_{r}(k) a^{(\nu)}_{\phi}(k) ) ]\nonumber\\
    &=& -2\hbar \mathrm{Im} [ a^{(\nu)*}_{r}(k) a^{(\nu)}_{\phi}(k) ]
\end{eqnarray}
and $\langle L_r\rangle_{\nu}(k)=0$ in the most generally coupled case. This means that the orbital texture lacks a radial component. This vanishing of $\langle \hat L_r\rangle_\nu(k)$ follows directly from the structure of Eq.~\ref{eq-generalBlochSpinor}: only the $p_r$ amplitude is imaginary, whereas the $p_\phi$ and $p_z$ amplitudes are real. Therefore $\mathrm{Im}\!\left[a_{\phi}^{(\nu)*}(k)a_{z}^{(\nu)}(k)\right]=0$, and the radial component of the orbital texture vanishes identically for the present single-helix model.

These expressions show directly that different components of the local orbital-angular-momentum texture arise from different inter-orbital coherences:
\begin{itemize}
\item $\langle \hat L_r \rangle_\nu(k)$ is controlled by the coherence between $p_\phi$ and $p_z$,
\item $\langle \hat L_\phi \rangle_\nu(k)$ is controlled by the coherence between $p_r$ and $p_z$,
\item $\langle \hat L_z \rangle_\nu(k)$ is controlled by the coherence between $p_r$ and $p_\phi$.
\end{itemize}

\begin{figure} [ht]
    \centering
    \includegraphics[width=1\linewidth]{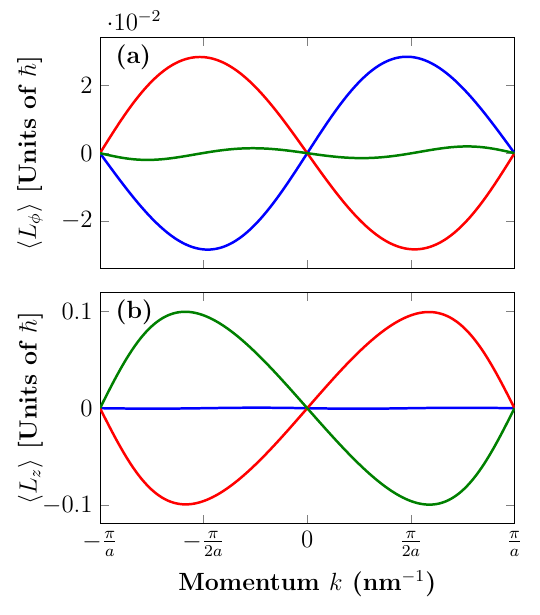}
    \caption{Exact local orbital angular momentum textures for the fully hybridized $3 \times 3$ helical Hamiltonian. (a) Azimuthal texture $\langle L_{\phi} \rangle$ generated primarily by the $p_r-p_z$ coherence. (b) Longitudinal texture $\langle L_z \rangle$ generated primarily by the $p_r-p_\phi$ coherence. The textures corresponding to the lowest energy band $E_1$ (online blue), the middle energy band $E_2$ (online red), and the highest energy band $E_3$ (online green) are shown. The texture for the radial angular moment is always zero within the single helix model presented here.}
    \label{fig:exact_textures}
\end{figure}

The structure of the Bloch Hamiltonian makes the origin of these components transparent. Since the $(z,\phi)$ sector is even in momentum, while the $(z,r)$ and $(r,\phi)$ sectors are odd in momentum, the orbital texture naturally decomposes into even and odd parts under $k\to -k$. 
 Therefore, the Bloch states generally carry a full local orbital-angular-momentum vector texture. Therefore, although the $(p_z,p_\phi)$ sector provides an even-in-$k$ coherence, it does not generate a surviving radial orbital texture in the present single-helix model. By contrast, the components $\langle \hat L_\phi\rangle_\nu(k)$ and $\langle \hat L_z\rangle_\nu(k)$ arise from the odd-in-$k$ hybridization channels enforced by the helical geometry. The textures for the full $3\times 3$ Hamiltonian are given in Fig.~\ref{fig:exact_textures}.

In the following section, we will show that, although the full local texture contains all three components in general, only the odd-in-momentum part contributes to the current-induced orbital response. In particular, within the present minimal helical model, the $(p_z,p_r)$ sector provides the dominant transport-active contribution to the orbital Edelstein.

\subsection{Specialization to the \(t_{rz}\)-only mixing sector}

It is useful to specialize the general three-orbital texture formulas to the case in which the only inter-orbital hopping retained is the \(p_r\)-\(p_z\) mixing, namely
$t_{r\phi}=0,~
t_{\phi z}=0,~
t_{rz}\neq 0.$
In that case, the Bloch Hamiltonian keeps the full three-orbital structure, but the \(p_\phi\) orbital decouples and is given by Eq.~\ref{eq-decoupling}.
 
-{\it The $p_{\phi}$ uncoupled band:},
\[
|u_\phi(k)\rangle=|p_\phi\rangle,
\qquad
E_\phi(k)=\xi_\phi(k),
\]
for which all orbital-texture components vanish:
\[
\langle L_r\rangle_\phi(k)=
\langle L_\phi\rangle_\phi(k)=
\langle L_z\rangle_\phi(k)=0.
\]

-{\it The $p_r-p_z$ sector $\pm$ bands:}

The remaining two bands arise from the \(p_r\)-\(p_z\) block
\[
H_{rz}(k)=
\begin{pmatrix}
A_k & iY_k\\
-iY_k & C_k
\end{pmatrix}.
\]
Defining
\begin{eqnarray}
d_0(k)&=&\frac{A_k+C_k}{2},
\nonumber\\
\delta(k)&=&\frac{A_k-C_k}{2},
\nonumber\\
\Omega(k)&=&\sqrt{\delta(k)^2+Y_k^2},
\end{eqnarray}
The two hybridized eigenvalues are
$E_\pm(k)=d_0(k)\pm \Omega(k)$.
A convenient parametrization of the corresponding normalized eigenvectors is
 \begin{align*}
|u_+(k)\rangle
&=
\cos\theta_k\,|p_r\rangle
-i\sin\theta_k\,|p_z\rangle,
\\[4pt]
|u_-(k)\rangle
&=
\sin{\theta_k}\,|p_r\rangle
+i\cos{\theta_k}\,|p_z\rangle,
\end{align*}
with
\[
\cos2\theta_k=\frac{\delta(k)}{\Omega(k)},
\qquad
\sin2\theta_k=\frac{Y_k}{\Omega(k)}.
\]
For these states, one has \(a_\phi^{(\pm)}(k)=0\), and therefore the general three-orbital expressions reduce immediately to
\[
\langle L_r\rangle_\pm(k)=0,
\qquad
\langle L_z\rangle_\pm(k)=0,
\]
while
\begin{align}
\langle L_\phi\rangle_\pm(k)
&=
-2\hbar\,\mathrm{Im}\!\left[a_z^{(\pm)*}(k)a_r^{(\pm)}(k)\right]
\nonumber\\[4pt]
&=
\mp \hbar\,\sin 2\theta_k
=
\mp \hbar\,\frac{Y_k}{\Omega(k)}.
\end{align}
Thus
\[
{
\langle L_\phi\rangle_\pm(k)
=
\mp \hbar\,
\frac{Y_k}
{\sqrt{\delta(k)^2+Y_k^2}}
}
\]
with
\[
{
\langle L_r\rangle_\pm(k)=0,
\qquad
\langle L_z\rangle_\pm(k)=0.
}
\]

\begin{figure}
    \centering
    \includegraphics[width=0.7\linewidth]{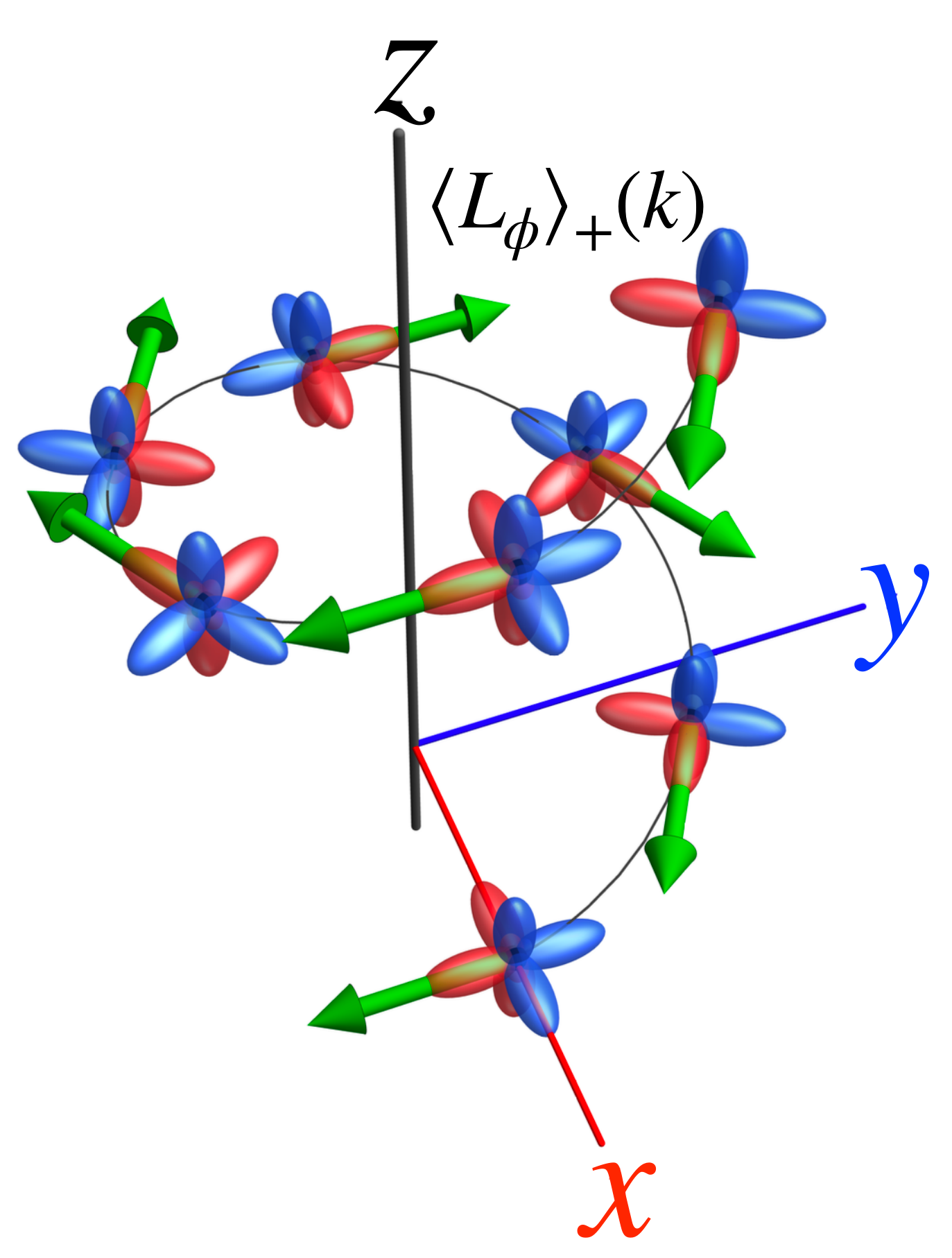}
    \caption{Orbital angular momentum textures given by Eq.\ref{Eq-OrbitalMomentumReduced}. for this case we only have a $\langle L_{\phi}\rangle(k)$ due to the coherence of $p_z$ and $p_r$ orbitals. The length of the orbital angular momentum vector (online green) depends on $k$ and is odd under $k\rightarrow -k$. }
    \label{fig:OrbitalTexture}
\end{figure}

\begin{figure} [ht]
    \centering
    \includegraphics[width=1\linewidth]{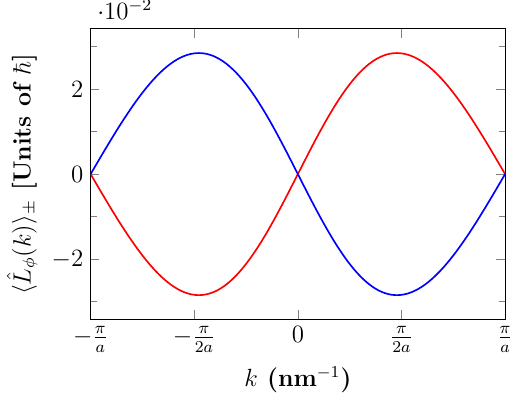}
    \caption{Azimuthal Orbital Angular Momentum Texture $\langle L_{\phi} \rangle$ generated by the ($p_z,p_r$) sector. The texture generated by $E_+$ (online red) has the same shape as the texture generated by $E_-$ (online blue), but with the opposite sign.}
    \label{fig:azimutal_texture2}
\end{figure}

Using \(Y_k=2t_{rz}\sin(ka)\), this becomes
\[
{
\langle L_\phi\rangle_\pm(k)
=
\mp \hbar\,
\frac{2t_{rz}\sin(ka)}
{\sqrt{\left[\frac{A_k-C_k}{2}\right]^2+4t_{rz}^2\sin^2(ka)}}
}
\]
or, if chirality is written explicitly,
\begin{equation}
\langle L_\phi\rangle_\pm(k)
=
\mp \hbar\,
\frac{2\eta\,t_{rz}\sin(ka)}
{\sqrt{\left[\frac{A_k-C_k}{2}\right]^2+4t_{rz}^2\sin^2(ka)}}.
\label{Eq-OrbitalMomentumReduced}
\end{equation}
The previous component is depicted in real space in Fig.~\ref{fig:OrbitalTexture} and in reciprocal space in Fig.~\ref{fig:azimutal_texture2}. This specialized limit makes clear the distinction between the full texture and the transport-active sector. The full three-orbital theory admits, in principle, the three components
$\bigl(\langle L_r\rangle,\langle L_\phi\rangle,\langle L_z\rangle\bigr)$,
but once only the \(p_r\)-\(p_z\) coherence is retained, the only nonzero component is $\langle L_\phi\rangle(k)$,
because \(L_r\) requires \((p_\phi,p_z)\) coherence and \(L_z\) requires \((p_r,p_\phi)\) coherence. Hence, the \(t_{rz}\)-only sector does not describe the full orbital texture; it isolates the single transport-active component selected by the surviving inter-orbital mixing. 

\section{Equilibrium Orbital currents}
 
We now compute the orbital currents, first considering the possibility of an equilibrium component resulting from the antisymmetry of the orbital texture. We work within the full three-orbital framework and explicitly track the vector and non-conserved character of the orbital angular momentum.

\subsection{Band structure and group velocity}

The Bloch Hamiltonian $\mathcal{H}(k)$ has three energy bands $E_\nu(k)$, defined by
\begin{equation}
\mathcal{H}(k)\, |u_\nu(k)\rangle = E_\nu(k)\, |u_\nu(k)\rangle,
\qquad \nu = 1,2,3.
\end{equation}
The group velocity in each band is
\begin{equation}
v_\nu(k) = \frac{1}{\hbar}\frac{\partial E_\nu(k)}{\partial k}.
\label{eq:velocity_full}
\end{equation}
Because the system preserves time-reversal symmetry, the band energies satisfy
$E_\nu(k) = E_\nu(-k)$,
so that
$v_\nu(-k) = -v_\nu(k)$.
The orbital angular momentum current associated with component $\alpha = r,\phi,z$ is defined as
\begin{equation}
\hat{J}^{(L_\alpha)}(k)
=
\frac{1}{2}
\left\{
\hat{v}(k), \hat{L}_\alpha
\right\},
\label{eq:current_operator_full}
\end{equation}
where $\hat{v}(k) = \frac{1}{\hbar}\partial_k \mathcal{H}(k)$.
Projecting onto the band eigenstates, the expectation value becomes
\begin{equation}
J^{(L_\alpha)}_\nu(k)
\simeq
v_\nu(k)\,
\langle \hat{L}_\alpha(k) \rangle_\nu,
\label{eq:current_expectation}
\end{equation}
which is valid in the adiabatic approximation, where we treat the Bloch electron as a wavepacket in a single band, carrying an internal orbital moment that is convected by the band velocity. Thus, the orbital current is determined by the product of the band velocity and the orbital-angular-momentum texture.

In equilibrium, the total orbital current is given by
\begin{equation}
J_{\rm eq}^{(L_{\alpha})}=\sum_\nu \int \frac{dk}{2\pi}
f_0(E_\nu(k))\, J^{(L_\alpha)}_\nu(k),
\end{equation}
because $v_\nu(k)$ is odd in $k$, while the equilibrium distribution satisfies $f_0(k)=f_0(-k)$, we have that the odd orbital texture would give an equilibrium persistent-like orbital current. This is allowed by the time-reversal symmetry we have in the problem so far.

Such an orbital current in the infinite chain is interesting because, in the presence of an interface, it would lead to a buildup of orbital angular momentum and magnetic moment. We argue this by resorting to the continuity equation for the orbital angular momentum
\begin{equation}
    \partial_t\langle L_{\phi}\rangle+\partial_z J^{L_{\phi}}=\cal T_{\phi},
\end{equation}
where $\cal T_{\phi}$ is a torque due to the ending interface or coupling to the lattice, giving it a twist\cite{ShiNiu2006}. In fact, this non-conservation of the angular momentum can then also result in an end accumulated angular momentum/magnetic moment that accounts for the finite $\partial_z J^{L_{\phi}}$ that we need to take the orbital current to zero at the boundary. Note that although the equilibrium $\langle L_{\phi}\rangle=0$ for the asymmetric dispersion, this is not essential for the above argument. The torque $\cal T_{\phi}=\cal T_{\phi}^{\rm static}+\cal T_{\phi}^{\rm diss}$ can be partially composed of a static polarization torque and another part that relaxes it into phonons, molecular deformation, substrate motion, etc. So some part of the torque may remain as a permanent magnetic moment, possibly related to the magnetic interaction seen in experiments between finite molecular helices or between helices and a magnetized substrate that separates a racemic mixture. The two components remaining of the end orbital magnetic moment will be in the $\phi$ and $z$ directions, fitting the qualitative scenarios in ref.\cite{NaamanRacemicSeparation}.  

\begin{figure}
    \centering
    \includegraphics[width=1.0\linewidth]{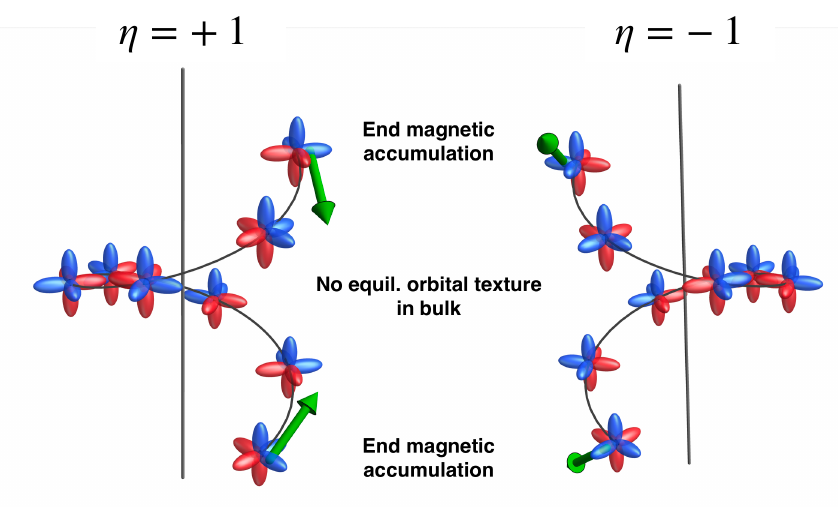}
    \caption{Accumulation of orbital angular momentum and therefore magnetic moment due to termination of a helix that sustains an equilibrium orbital current (for both chiralities). Both surviving components ($\hat\phi$, $z$) are represented. No orbital texture is present in the bulk once we sum over all bands and the distribution of occupied states. Note the opposite magnetizations that would select chirality with the interaction to a magnetized surface. The $\phi$ component gets averaged out (see text) while the $z$ component remains.}
    \label{fig:EquilAccum}
\end{figure}

A rough estimate of the resulting magnetic moment's magnetization depends on whether we are talking about $J^{L_{\phi}}$ or $J^{L_{z}}$ and the details of how many sites it takes for the gradient of the orbital current to decay toward zero. For $J^{L_{\phi}}$, dependent on $t_{rz}$, as $\hat\phi$ is a local vector, then the changes in the direction are vector sums and tend to cancel, but a finite moment may remain. On the other hand, for $J^{L_{z}}$ dependent on $t_{r\phi}$, $\hat z$ is fixed, and no vector averaging ensues.  We can estimate an upper bound for the orbital moment, the $z$ component: At site $n$ toward the end of the helix as
\begin{equation}
    |{\mathbf \mu}_{L,n}|\sim-\mu_B\frac{|\langle L_{z}\rangle|}{\hbar},
\end{equation}
where $\mu_B$ is the Bohr magneton. If $|\langle L_{z}\rangle|\sim 0.01\hbar-0.1\hbar$ then $\mu_{end}=\mu_B\sum_{n\in{\rm end}}\langle L_{z}\rangle/\hbar\sim 0.15\mu_B$ for a three-site gradient to taper off the orbital momentum current.
The interaction between finite helices and a magnetic substrate could then be used to verify this orbital contribution (see figure\ref{fig:EquilAccum}). Note that although we have mostly discussed $J^{L_{\phi}}$, the same arguments apply to $J^{L_z}$, which is also odd under $k$ inversion.

When we analyze the equilibrium orbital angular momentum texture, on the other hand
\begin{equation}
L_{\alpha,\rm eq}=\sum_\nu \int \frac{dk}{2\pi}
f_0(E_\nu(k))\, \langle L_{\alpha}\rangle_{\nu}(k),
\end{equation}
the equilibrium texture is zero as there are no even in $k$ contributions to the texture (as depicted in Fig.~\ref{fig:EquilAccum}).

\section{Edelstein Orbital textures and currents}

We now turn to the linear response induced by a longitudinal electric
field $E_{\parallel}$ applied along the helical axis. Within the
relaxation-time approximation of the distribution
function acquires the correction
\begin{equation}
\delta f_{\nu}(k)=
-eE_{\parallel}\tau\,v_{\nu}(k)
\left(-\frac{\partial f_{0}}{\partial E}\right)_{E=E_{\nu}(k)} .
\end{equation}
where $\tau$ is the relaxation time, not determined in our approach. Since $E_{\nu}(-k)=E_{\nu}(k)$ in the nonmagnetic problem, one has
$v_{\nu}(-k)=-v_{\nu}(k)$, and the derivative of the distribution function is even under the change of sign of $k$, $\delta f_{\nu}(k)$ is odd in $k$.
The induced orbital angular-momentum density is
\begin{equation}
\delta\langle L_{\alpha}\rangle =
\sum_{\nu}\int_{\mathrm{BZ}}\frac{dk}{2\pi}\,
\langle \hat L_{\alpha}\rangle_{\nu}(k)\,\delta f_{\nu}(k),
\qquad \alpha=r,\phi,z .
\end{equation}
This defines the orbital Edelstein susceptibility through $\delta\langle L_{\alpha}\rangle=\chi_{L_{\alpha}E_{\parallel}}$ where
\begin{eqnarray}
\chi_{L_{\alpha}}=
-e\tau \sum_{\nu}\int_{\mathrm{BZ}}\frac{dk}{2\pi}\,
v_{\nu}(k)\,\langle \hat L_{\alpha}\rangle_{\nu}(k)
\left(-\frac{\partial f_{0}}{\partial E}\right)_{E=E_{\nu}(k)}.\nonumber\\
\end{eqnarray}
To make the parity structure explicit, we decompose the texture into even and
odd parts,
$\langle \hat L_{\alpha}\rangle_{\nu}(k)=
\langle \hat L_{\alpha}\rangle^{\mathrm{even}}_{\nu}(k)+
\langle \hat L_{\alpha}\rangle^{\mathrm{odd}}_{\nu}(k)$.
 Only the odd-in-$k$ part of the
orbital texture contributes to the linear Edelstein response:
\begin{equation}
\chi_{L_{\alpha}}=
-e\tau \sum_{\nu}\int_{\mathrm{BZ}}\frac{dk}{2\pi}\,
v_{\nu}(k)\,
\langle \hat L_{\alpha}\rangle^{\mathrm{odd}}_{\nu}(k)
\left(-\frac{\partial f_{0}}{\partial E}\right)_{E=E_{\nu}(k)} .
\end{equation}
The two transport-active odd sectors of the present model are $\langle L_{\phi}\rangle$ and $\langle L_{z}\rangle$ (the $\langle L_r\rangle$ being always zero by symmetry for a single helix).
The odd orbital texture in $ k$-space then generates a finite net orbital texture upon averaging over the three bands and the BZ. This contrasts with the equilibrium orbital texture that was zero in the previous section.
On the other hand, for the induced orbital current, we proceed as in the previous section on equilibrium currents. Naming
 the exact band expectation value as $j_{\alpha,\nu}(k)=
\langle u_{\nu}(k)|\hat J^{(L_{\alpha})}(k)|u_{\nu}(k)\rangle$
The field-induced orbital current is
\begin{equation}
\delta\langle J^{(L_{\alpha})}\rangle=
\sum_{\nu}\int_{\mathrm{BZ}}\frac{dk}{2\pi}\,
j_{\alpha,\nu}(k)\,\delta f_{\nu}(k)
\equiv \sigma_{L_{\alpha}} E_{\parallel},
\end{equation}
where $\alpha=\phi,z$, with
\begin{equation}
\sigma_{L_{\alpha}}=
-e\tau \sum_{\nu}\int_{\mathrm{BZ}}\frac{dk}{2\pi}\,
j_{\alpha,\nu}(k)\,v_{\nu}(k)
\left(-\frac{\partial f_{0}}{\partial E}\right)_{E=E_{\nu}(k)} .
\end{equation}
If one keeps only the projected intraband contribution, it reduces to $
j_{\alpha,\nu}(k)\simeq
v_{\nu}(k)\,\langle \hat L_{\alpha}\rangle_{\nu}(k)$, and therefore
\begin{equation}
\sigma^{\mathrm{intra}}_{L_{\alpha}}=
-e\tau \sum_{\nu}\int_{\mathrm{BZ}}\frac{dk}{2\pi}\,
v_{\nu}^{2}(k)\,\langle \hat L_{\alpha}\rangle_{\nu}(k)
\left(-\frac{\partial f_{0}}{\partial E}\right)_{E=E_{\nu}(k)}.
\end{equation}
In this projected expression, the factor $v_{\nu}^{2}(k)$ is even in $k$, so the
linear orbital-current conductivity probes the even-in-$k$ part of the orbital texture. We then come to the very strong conclusion that for the linear response of a single helix, all the components of the {\it the projected linear orbital-current conductivities vanish}.

 Thus, in the linear regime, the present model supports a finite orbital
Edelstein texture but no projected longitudinal orbital-current conductivity. Finite conductivity will not be achieved by including the full anticommutator for the current (due to parity), but the finite texture in $k$-space could be restored in the asymmetric double-helix model, which can produce even in $k$ dispersion. We leave this case for future work.

The physical content of the two active odd sectors may therefore be summarized
as
\begin{equation}
t_{rz}\neq 0
\quad\Longrightarrow\quad
\langle \hat L_{\phi}\rangle^{\mathrm{odd}}(k)\neq 0
\quad\Longrightarrow\quad
\chi_{L_{\phi}}\neq 0,
\end{equation}
\begin{equation}
t_{\phi r}\neq 0
\quad\Longrightarrow\quad
\langle \hat L_{z}\rangle^{\mathrm{odd}}(k)\neq 0
\quad\Longrightarrow\quad
\chi_{L_{z}}\neq 0.
\end{equation}
In both cases, the chirality of the helix fixes the sign of the response through
the sign of the odd inter-orbital hybridization.
\begin{equation}
\eta\rightarrow -\eta
\qquad\Longrightarrow\qquad
\chi_{L_{\phi,z}}\rightarrow -\chi_{L_{\phi,z}} .
\end{equation}

\begin{figure}[h]
    \centering
    \includegraphics[width=1\linewidth]{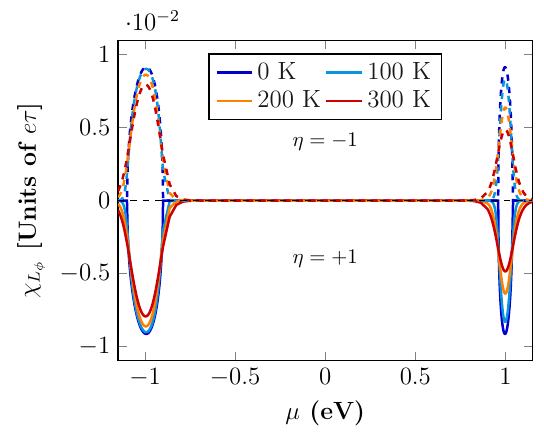}
    \caption{Temperature dependence of the azimuthal orbital Edelstein susceptibility, $\chi_{L_\phi}$, as a function of the chemical potential $\mu$. The response originates from the odd-in-momentum hybridization within the $(p_r, p_z)$ orbital sector. The solid lines represent the response for positive quirality ($\eta=+1$) and the dashed lines for negative quirality ($\eta=-1$). At absolute zero ($T = 0$ K, online dark blue line), the accumulation is strictly constrained to the narrow energy bands, exhibiting sharp van Hove singularities at the band edges separated by a massive $\sim 1.9$ eV zero-response gap where no electronic states exist. Because the kinetic bandwidths ($\sim 80$--$200$ meV) are small compared to the on-site energy difference, thermal fluctuations ($k_B T$) become significant at relatively low temperatures.}
    \label{fig:chi_phi}
\end{figure}
Figure \ref{fig:chi_phi} illustrates the chemical potential ($\mu$) and temperature dependence of the azimuthal orbital Edelstein susceptibility, $\chi_{L_\phi}$. In contrast to weakly gapped configurations, this parameter regime enforces a massive zero-response gap spanning approximately $1.9$ eV. When $\mu$ resides within this void—roughly between $-0.90$ eV and $0.96$ eV—the strict absence of electronic states at the Fermi surface yields a universally null orbital accumulation. At $T = 0$K, the response is tightly confined to the ultra-narrow kinetic-energy bands, manifesting as sharp van Hove singularities at the band edges, where new momentum states abruptly enter the integration domain. Because these kinetic bandwidths ($\sim 80$ to $200$ meV) are extremely narrow relative to the on-site gap, the system is highly sensitive to thermal fluctuations ($k_B T$). At finite temperatures, particularly at $300$ K, thermal smearing heavily washes out the band-edge singularities. This broadens the active response window but severely suppresses the peak accumulation magnitude as electrons are thermally excited across the entire band. Furthermore, the macroscopic susceptibility remains exclusively positive across all active energy ranges. This unipolar behavior is a direct consequence of strict parity cancellation within the isolated parabolic bands. The product of the group velocity ($v_k$) and the chiral texture ($\langle L_\phi \rangle_k$) maintains a consistent positive sign across the Brillouin zone, which gets multiplied with the negative electronic charge prefactor inherent to the Edelstein response, ensuring unidirectional orbital accumulation regardless of the occupied band.

\section{Spin Edelstein response: bare spin--orbit coupling versus orbital transduction}

The orbital Edelstein mechanism derived in Sec.~IV produces a nonequilibrium orbital texture without requiring atomic spin--orbit coupling. We now compare the spin polarization obtained from a conventional spin Edelstein mechanism with that generated when the odd orbital texture of the helix is converted into spin polarization by a local spin--orbit interaction. In the present three-orbital model, the relevant transduction channel is the longitudinal orbital texture $\langle \hat L_z\rangle$, generated by the odd $(p_r,p_\phi)$ sector. We therefore do not use the radial texture $\langle \hat L_r\rangle$, which vanishes for the single helix considered here.

  Throughout this section we use the same relaxation-time approximation introduced in Sec.~IV. For any observable $\hat O$, the field-induced correction is
\begin{equation}
\delta\langle O\rangle
=
\sum_{\nu}\int_{\mathrm{BZ}}\frac{dk}{2\pi}\,
\langle \hat O\rangle_{\nu}(k)\,\delta f_{\nu}(k).
\end{equation}

We first consider, as a reference, a single-orbital chain with spin-orbit coupling but no orbital channel. A minimal time-reversal-symmetric Bloch Hamiltonian is
\begin{equation}
H_{\mathrm{SOC}}(k)=\varepsilon_0(k)\,\mathds{1}_s+b(k)\sigma_z,
\qquad
b(-k)=-b(k).
\end{equation}
Its eigenenergies are
\begin{equation}
E_{\pm}(k)=\varepsilon_0(k)\pm |b(k)|,
\end{equation}
and the spin expectation values are
\begin{equation}
\langle S_z\rangle_{\pm}(k)=\pm \frac{\hbar}{2}\,\mathrm{sgn}[b(k)].
\end{equation}
At zero temperature, and for weak spin splitting, the corresponding one-dimensional spin Edelstein response reduces to
\begin{equation}
\delta S_z^{(\mathrm{SOC})}
\simeq
-\frac{eE_{\parallel}\tau}{\pi\hbar}\,b(k_F).
\end{equation}

We now turn to the orbital mechanism. To isolate the part of the helical Hamiltonian that carries the $\langle \hat L_z\rangle$ texture, we project the full Bloch Hamiltonian onto the $(p_r,p_\phi)$ subspace. The resulting $2\times 2$ block is
\begin{equation}
H_{r\phi}(k)
=
d^{(z)}_0(k)\tau_0+\delta_z(k)\tau_z-X_k\tau_y,
\end{equation}
with
\begin{eqnarray}
d^{(z)}_0(k)&=&\frac{A_k+B_k}{2},\nonumber\\
\delta_z(k)&=&\frac{A_k-B_k}{2},\nonumber\\
X_k&=&2t_{r\phi}\sin(ka),
\end{eqnarray}
and
\begin{equation}
\Omega_z(k)=\sqrt{\delta_z(k)^2+X_k^2}.
\end{equation}
We use the sub- and superscripts $z$ to emphasize we are only considering this sector, the functions $\delta(k)$ and $d_0(k)$ are defined as before.

The two eigenvalues of this block are
\begin{equation}
E^{(z)}_{\pm}(k)=d^{(z)}_0(k)\pm \Omega_z(k).
\end{equation}
In the $(p_r,p_\phi)$ basis the local orbital angular momentum operator is
\begin{equation}
\hat L_z=-\hbar \tau_y.
\end{equation}
Therefore, the projected band textures are
\begin{equation}
\langle \hat L_z\rangle_{\pm}(k)
=
\pm \hbar\,\frac{X_k}{\Omega_z(k)}.
\end{equation}
Since $X_k=2t_{r\phi}\sin(ka)$ is odd in momentum, the texture $\langle \hat L_z\rangle_{\pm}(k)$ is also odd under $k\to -k$, exactly as required for a finite orbital Edelstein response.

To describe orbital-to-spin transduction in the molecular region, we introduce an effective local spin--orbit interaction
\begin{equation}
H_{LS}^{(z)}
=
\frac{\lambda_z}{\hbar^2}\,\hat L_z \hat S_z
=
\frac{\lambda_z}{2\hbar}\,\hat L_z \sigma_z
=
-\frac{\lambda_z}{2}\,\tau_y\sigma_z,
\end{equation}
where $\lambda_z$ is an effective orbital-to-spin conversion energy scale. Projecting onto the orbital band $\nu=\pm$ gives
\begin{equation}
\langle u_\nu(k)|H_{LS}^{(z)}|u_\nu(k)\rangle
=
b_{\mathrm{eff}}^{(\nu)}(k)\,\sigma_z,
\end{equation}
with the induced effective spin splitting
\begin{equation}
b_{\mathrm{eff}}^{(\nu)}(k)
=
\frac{\lambda_z}{2\hbar}\,
\langle \hat L_z\rangle_{\nu}(k)
=
\nu\,\frac{\lambda_z}{2}\,\frac{X_k}{\Omega_z(k)}.
\end{equation}
This effective field is odd in momentum and therefore produces a spin Edelstein response.

In the weak-transduction limit, the resulting nonequilibrium spin polarization is
\begin{equation}
\delta S_z^{(\mathrm{helix}+LS)}
\simeq
-\frac{eE_{\parallel}\tau}{2\pi\hbar}
\sum_{\nu=\pm}\sum_{k_{F,\nu}}
b_{\mathrm{eff}}^{(\nu)}(k_{F,\nu})\,
\mathrm{sgn}\!\left[v_{\nu}(k_{F,\nu})\right].
\end{equation}
Substituting the projected splitting yields
\begin{equation}
\delta S_z^{(\mathrm{helix}+LS)}
\simeq
-\frac{eE_{\parallel}\tau\,\lambda_z}{4\pi\hbar}
\sum_{\nu=\pm}\sum_{k_{F,\nu}}
\nu\,
\frac{X_{k_{F,\nu}}}{\Omega_z(k_{F,\nu})}\,
\mathrm{sgn}\!\left[v_{\nu}(k_{F,\nu})\right].
\end{equation}
Using $X_k=2t_{r\phi}\sin(ka)$, one sees explicitly that the induced spin polarization changes sign with chirality:
\begin{equation}
t_{r\phi}\rightarrow -t_{r\phi}
\qquad\Longrightarrow\qquad
\delta S_z^{(\mathrm{helix}+LS)}
\rightarrow
-\delta S_z^{(\mathrm{helix}+LS)}.
\end{equation}

A simple comparison with the bare spin--orbit mechanism follows from the two preceding expressions. Parametrically,
\begin{equation}
\frac{\delta S_z^{(\mathrm{helix}+LS)}}{\delta S_z^{(\mathrm{SOC})}}
\sim
\frac{\lambda_z}{2\alpha}\,
\frac{X(k_F)}{\Omega_z(k_F)\sin(k_Fa)}
=
\frac{\lambda_z t_{r\phi}}{\alpha\,\Omega_z(k_F)},
\end{equation}
up to filling-dependent factors associated with the number of Fermi points. The orbital route therefore replaces the small relativistic scale $\alpha$ by the geometrically generated odd hybridization $t_{r\phi}$, multiplied by the conversion scale $\lambda_z$.

\section{Spin accumulation versus spin chemical potentials}

The mechanism described above generates a nonequilibrium spin density,
\begin{equation}
\delta S_z\neq 0,
\end{equation}
after orbital-to-spin conversion. In the coherent molecular region, this quantity should be interpreted as a spin-dependent nonequilibrium occupation,
\begin{equation}
\delta f_{\uparrow}(k)\neq \delta f_{\downarrow}(k),
\end{equation}
rather than as distinct local spin chemical potentials.

The language of spin-dependent chemical potentials, $\mu_{\uparrow}-\mu_{\downarrow}$, becomes appropriate only after rapid momentum and energy relaxation allow each spin channel to equilibrate separately. This regime is expected in the metallic electrodes rather than inside the coherent molecular segment. The role of the helix is therefore to inject a nonequilibrium spin imbalance into the contacts, where it may then be described in the usual spin-injection language.

The orbital route replaces the small relativistic scale $\alpha$ by the geometrically generated odd hybridization $t_{r\phi}$, multiplied by the conversion scale $\lambda_z$. Since $t_{r\phi}$ originates from the helical geometry rather than from atomic relativistic effects, substantial enhancement of spin generation can occur even when the intrinsic spin--orbit interaction is weak.

An alternative possibility is that the orbital angular momentum generated within the molecule is predominantly converted into spin at the contacts, where spin-orbit coupling may be stronger.\cite{Matityahu2016Dubi} This mechanism has often been invoked in the interpretation of CISS experiments. However, experimental observations indicating only a weak dependence of the signal on the contact metal suggest that interfacial conversion may not, by itself, account for the full effect.\cite{reviewPaltiel} This motivates the bulk transduction scenario considered here, in which the helical structure first generates an orbital Edelstein texture and the local molecular spin--orbit interaction then converts it into spin polarization.

\section{Summary and Conclusions}

Understanding the origin of spin polarization in chiral molecular conductors remains a central problem in chirality-induced spin selectivity, particularly because large spin signals are observed in systems with weak intrinsic spin--orbit coupling. In this work, we introduced a minimal tight-binding description showing that orbital angular momentum textures and currents arise intrinsically in a one-dimensional chiral structure, without requiring atomic spin--orbit interaction. Using a three-orbital per-site Hamiltonian in the local basis $(p_r,p_\phi,p_z)$, we showed that the lack of inversion-related cancellation in the helical geometry generates odd-in-momentum inter-orbital hybridization terms, which in turn produce a momentum-dependent orbital angular momentum texture in the Bloch states.

A first structural result of the single-helix model is that the radial orbital-angular-momentum texture vanishes identically, whereas the azimuthal and longitudinal components remain finite. The surviving $\langle L_\phi\rangle$ and $\langle L_z\rangle$ textures arise from the odd $(p_z,p_r)$ and $(p_r,p_\phi)$ sectors, respectively, and therefore constitute the only transport-active channels of the present geometry.

A second central result is the contrast between equilibrium and nonequilibrium behavior. In equilibrium, the average orbital texture vanishes by parity, yet persistent-like orbital angular momentum currents may still exist because the band velocity is odd in momentum and multiplies an odd orbital texture. For a finite helix, the interruption of such an equilibrium current at the molecular terminations implies a chirality-dependent accumulation of orbital magnetic moment at the ends, with the dominant surviving contribution expected along the molecular axis.

When a nonequilibrium correction $\delta f$ is generated by an external electric field or bias, one can define response functions for both the induced average texture and the orbital angular momentum current. In the linear Edelstein regime, the present single-helix model supports a finite orbital susceptibility $\chi_{L\alpha}$ but no projected longitudinal orbital-current conductivity $\sigma_{L\alpha}$. The reason is purely one of parity: the projected linear current response contains the even factor $v_\nu^2(k)$, so it probes the even-in-$k$ part of the orbital texture, while the surviving texture channels of the single helix are odd in momentum. Thus, the leading linear nonequilibrium effect of the single helix is an induced orbital texture rather than a projected longitudinal orbital current. This conclusion, however, is specific to the linear regime. Beyond linear response, higher-order corrections to $\delta f$ do not in general preserve the same parity structure, and orbital-current contributions are therefore no longer excluded by parity arguments alone. Furthermore, this conclusion is for the single helix, a double-helix geometry provides a natural route to a nonzero longitudinal
orbital-current conductivity. The additional strand degree of freedom and the
inter-strand couplings generally destroy the special real-gauge structure of the
single-chain Hamiltonian, allowing a finite radial orbital texture $L_r(k)$ to
develop. When this texture contains an even-in-$k$ component, it produces a
nonzero orbital conductivity $\sigma_{L_r}$ in linear response. This we leave for future work.

When spin degrees of freedom are included, the orbital Edelstein texture acts as a source of spin polarization through orbital-to-spin transduction. In the present formulation, the relevant channel is the longitudinal orbital texture $\langle L_z\rangle$, which produces an effective spin splitting proportional to the orbital angular momentum expectation value. The important point is that the strength of the induced spin polarization is controlled by geometrically generated orbital overlaps, such as $t_{r\phi}$, multiplied by the orbital-to-spin conversion scale $\lambda_z$, rather than by the much smaller bare relativistic scale entering the conventional spin Edelstein effect. Spin generation mediated by orbital angular momentum can therefore be substantially stronger than the direct spin--orbit route. Experimental observations indicating only a weak dependence of the signal on the contact metal also suggest that interfacial conversion may not, by itself, account for the full effect, thereby favoring the bulk transduction scenario considered here.\cite{Matityahu2016Dubi,reviewPaltiel} An interesting prospect gleaned from the previous results is to measure the the average textures in crystaline uniaxial crystals of chiral strands, such as Tellurium\cite{calavalle2022gate}, or crystals with point chiral centers like Dichalcogenides\cite{inui2020chirality} or distorted perovskites\cite{yuntianzutic2025} could show surface magnetization due to an applied electric field by way of, e.g., local Kerr rotation measurements.

The physical picture that emerges is that chirality in a one-dimensional helix generates orbital angular momentum texture as the primary response, while spin polarization appears as a secondary consequence of orbital-to-spin transduction. In equilibrium, this can sustain persistent-like orbital currents whose interruption at the boundaries yields end magnetization. Under linear nonequilibrium driving, the same chirality generates a finite orbital Edelstein texture, while the projected longitudinal orbital-current conductivity remains forbidden by parity in the single helix. The single helix should therefore be viewed as the minimal and most constrained realization of the mechanism, while more complex geometries, such as asymmetric double helices, provide a natural route to restoring orbital-current channels that are absent in the simplest case. Such a nonequilibrium pathway can nevertheless coexist with previously proposed equilibrium mechanisms.\cite{mena2024Minimal,Bedoya2026}.

\section*{Acknowledgments}

We acknowledge financial support from the Universidad San Francisco de Quito through Poligrant POLI~41981 and the Université de Lorraine for an invited professor position (EM) where this work started. We dedicate this work to the memory of Vladimiro Mujica, whose insight and enthusiasm first drew us to CISS. We also acknowledge the use of AI-assisted tools for language refinement and discussion of theoretical directions during the preparation of this manuscript.

\vspace{1.5\baselineskip}
\vspace{1.5\baselineskip}

\appendix

\appendix

\section{Numerical values of the Tight-binding parameters}
\label{sec:parameters}

The representative on-site energies and hopping amplitudes \cite{hawke2010electronic} used in all numerical results in the paper are listed in Table~\ref{tab:tb_parameters}.

\begin{table}[htpb]
\centering
\caption{Tight-binding parameters\cite{hawke2010electronic} utilized for the calculation of the reduced band structure and orbital angular momentum textures.}
\label{tab:tb_parameters}
\begin{ruledtabular}
\begin{tabular}{lcc}
\textbf{Parameter} & \textbf{Symbol} & \textbf{Value} \\
\colrule
Lattice constant & $a$ & $0.34$ nm \\
Radial on-site energy & $\epsilon_r$ & $1.0$ eV \\
Azimuthal on-site energy & $\epsilon_\phi$ & $1.15$ eV \\
Longitudinal on-site energy & $\epsilon_z$ & $-1.0$ eV \\
Radial-radial hopping & $t_{rr}$ & $-0.020$ eV \\
Azimuthal-azimuthal hopping & $t_{\phi\phi}$ & $0.010$ eV \\
Longitudinal-longitudinal hopping & $t_{zz}$ & $0.050$ eV \\
Radial-longitudinal helical mixing & $t_{rz}$ & $-0.0143$ eV \\
Radial-azimuthal helical mixing & $t_{r\phi}$ & $0.0035$ eV \\
Azimuthal-longitudinal helical mixing & $t_{\phi z}$ & $0.0439$ eV 
\end{tabular}
\end{ruledtabular}
\end{table}

The tight-binding hopping parameters presented in Table \ref{tab:tb_parameters} are strictly derived from the underlying molecular geometry using the Slater-Koster projection rules defined in Eq. \ref{sk_parameters}. To physically constrain the model, we use fundamental orbital overlap integrals characteristic of $p$-orbital interactions between adjacent organic bases. Specifically, we set the strong head-to-head $\sigma$-bond integral to $V_{pp\sigma} = 80$ meV and the weaker parallel $\pi$-bond integral to $V_{pp\pi} = -21$ meV.

By projecting these chemical bond strengths onto the local cylindrical coordinate system of a right-handed B-DNA structure—characterized by an azimuthal twist angle of $\varphi = 36^\circ$ \cite{hawke2010electronic} between sequential sites—the phenomenological hopping amplitudes naturally emerge. The macroscopic structural anisotropy is entirely encapsulated by the spatial directional cosines ($\alpha, \beta, \gamma$), which are forced to obey the strict 3D geometric normalization condition $\alpha^2 + \beta^2 + \gamma^2 = 1$. The energetic difference between the fundamental bonds, defined as $\Delta \equiv V_{pp\sigma} - V_{pp\pi} = 101$ meV, acts as the primary driving amplitude for the off-diagonal chiral mixing. Consequently, this rigorous geometric framework ensures that the resulting Hamiltonian intrinsically captures the dominant longitudinal transport channel ($t_{zz}$) while correctly generating the asymmetric, symmetry-breaking helical couplings ($t_{rz}, t_{\phi z}, t_{r\phi}$) requisite for the orbital Edelstein response.

\section{Derivation of the effective spin splitting $b_{\mathrm{eff}}^{(\nu)}(k)$}

In this Appendix, we derive the effective spin splitting arising from the coupling between the helical orbital texture and spin. The derivation is written so as to match the main text, where the relevant transduction channel is the longitudinal orbital texture $\langle \hat L_z\rangle$ generated by the odd $(p_r,p_\phi)$ sector.

\subsection{Projected orbital Hamiltonian and local spin--orbit coupling}

To isolate the part of the helical Hamiltonian that carries the $\langle \hat L_z\rangle$ texture, we project the full three-orbital Bloch Hamiltonian onto the $(p_r,p_\phi)$ subspace. The resulting orbital Hamiltonian is
\begin{equation}
H_{r\phi}(k)=d_0^{(z)}(k)\tau_0+\delta_z(k)\tau_z-X_k\tau_y,
\end{equation}
defined in the main text.
 The corresponding orbital eigenvalues are
\begin{equation}
E_\nu^{(z)}(k)=d_0^{(z)}(k)+\nu\,\Omega_z(k),
\qquad \nu=\pm.
\end{equation}

In the $(p_r,p_\phi)$ basis, the local orbital angular momentum operator is
\begin{equation}
\hat L_z=-\hbar \tau_y.
\end{equation}
Therefore, in the unperturbed orbital eigenstates,
\begin{equation}
\langle u_\nu(k)|\tau_y|u_\nu(k)\rangle
=
-\nu\,\frac{X_k}{\Omega_z(k)},
\end{equation}
and hence
\begin{equation}
\langle \hat L_z\rangle_\nu(k)
=
-\hbar \langle u_\nu(k)|\tau_y|u_\nu(k)\rangle
=
\nu\,\hbar\,\frac{X_k}{\Omega_z(k)}.
\end{equation}

To describe orbital-to-spin transduction along the molecular axis, we add the effective local spin--orbit interaction
\begin{equation}
H_{LS}^{(z)}
=
\frac{\lambda_z}{\hbar^2}\hat L_z \hat S_z
=
\frac{\lambda_z}{2\hbar}\hat L_z \sigma_z
=
-\frac{\lambda_z}{2}\tau_y \sigma_z,
\end{equation}
where $\sigma_z$ acts in spin space and $\lambda_z$ is the effective orbital-to-spin conversion scale.

The full projected Hamiltonian is therefore
\begin{equation}
H(k)=H_{r\phi}(k)\otimes \mathds{1}_s + H_{LS}^{(z)}.
\end{equation}

\subsection{Exact spin-resolved spectrum}

Because the Hamiltonian contains only $\sigma_z$, it is block diagonal in the spin basis. For spin up and spin down, one obtains
\begin{equation}
H_{\uparrow}(k)
=
d_0^{(z)}(k)\tau_0+\delta_z(k)\tau_z-\left(X_k+\frac{\lambda_z}{2}\right)\tau_y,
\end{equation}
\begin{equation}
H_{\downarrow}(k)
=
d_0^{(z)}(k)\tau_0+\delta_z(k)\tau_z-\left(X_k-\frac{\lambda_z}{2}\right)\tau_y.
\end{equation}
The corresponding exact eigenvalues are
\begin{equation}
E_{\nu,\uparrow}(k)
=
d_0^{(z)}(k)
+
\nu\sqrt{\delta_z(k)^2+\left(X_k+\frac{\lambda_z}{2}\right)^2},
\end{equation}
\begin{equation}
E_{\nu,\downarrow}(k)
=
d_0^{(z)}(k)
+
\nu\sqrt{\delta_z(k)^2+\left(X_k-\frac{\lambda_z}{2}\right)^2},
\end{equation}
with $\nu=\pm$. Thus, the exact spin splitting within orbital band $\nu$ is
\begin{equation}
\Delta E_\nu(k)=E_{\nu,\uparrow}(k)-E_{\nu,\downarrow}(k).
\end{equation}

\subsection{Weak-transduction limit and effective field}

To connect with the effective description used in the main text, we assume that the transduction scale is small compared with the orbital splitting,
\begin{equation}
\left|\frac{\lambda_z}{2}\right| \ll \Omega_z(k).
\end{equation}
Expanding the square roots to first order in $\lambda_z$ gives
\begin{equation}
\sqrt{\delta_z(k)^2+\left(X_k\pm \frac{\lambda_z}{2}\right)^2}
\approx
\Omega_z(k)\pm \frac{\lambda_z}{2}\frac{X_k}{\Omega_z(k)}.
\end{equation}
Therefore,
\begin{equation}
E_{\nu,\uparrow}(k)
\approx
d_0^{(z)}(k)+\nu \Omega_z(k)+\nu \frac{\lambda_z}{2}\frac{X_k}{\Omega_z(k)},
\end{equation}
\begin{equation}
E_{\nu,\downarrow}(k)
\approx
d_0^{(z)}(k)+\nu \Omega_z(k)-\nu \frac{\lambda_z}{2}\frac{X_k}{\Omega_z(k)}.
\end{equation}

Comparing with the standard form
\begin{equation}
E_{\nu,\uparrow/\downarrow}(k)=E_\nu^{(z)}(k)\pm b_{\mathrm{eff}}^{(\nu)}(k),
\end{equation}
one identifies the effective spin splitting as
\begin{equation}
b_{\mathrm{eff}}^{(\nu)}(k)
=
\nu\,\frac{\lambda_z}{2}\frac{X_k}{\Omega_z(k)}.
\end{equation}

Using the orbital texture derived above, this may be written in the compact form
\begin{equation}
b_{\mathrm{eff}}^{(\nu)}(k)
=
\frac{\lambda_z}{2\hbar}\,
\langle \hat L_z\rangle_\nu(k).
\end{equation}
This is the result quoted in the main text.

Since $X_k=2t_{r\phi}\sin(ka)$ is odd in momentum, both $\langle \hat L_z\rangle_\nu(k)$ and $b_{\mathrm{eff}}^{(\nu)}(k)$ are odd under $k\to -k$. This is precisely the parity structure required for a finite spin Edelstein response generated by orbital-to-spin transduction.

The projected description above is valid as long as $\lambda_z$ does not strongly modify the orbital eigenstates themselves. When $\lambda_z$ becomes comparable to the orbital splitting $2\Omega_z(k)$, the orbital and spin sectors are strongly mixed, and the exact spectrum must be used instead of the linearized form.

\newpage
\bibliography{bio2.bib}

@article{Bedoya2026,
    author = {Bedoya, Valeria and Pastawski, Horacio M. and Fernández-Alcázar, Lucas J. and Medina, Ernesto},
    title = {Spin–charge transport in chirally induced spin selectivity},
    journal = {The Journal of Chemical Physics},
    volume = {164},
    number = {4},
    pages = {044306},
    year = {2026},
    month = {01},
    issn = {0021-9606},
    doi = {10.1063/5.0312418},
    url = {https://doi.org/10.1063/5.0312418}
}

@article{yuntianzutic2025,
author = {Liu, Yuntian and Shrestha, Reshna and Denisov, Konstantin and Ayala, Denzel and van Schilfgaarde, Mark and Nie, Wanyi and Žutić, Igor},
title = {Unconventional Spintronics from Chiral Perovskites},
journal = {Advanced Functional Materials},
volume = {35},
number = {52},
pages = {e09127},
doi = {https://doi.org/10.1002/adfm.202509127},
url = {https://advanced.onlinelibrary.wiley.com/doi/abs/10.1002/adfm.202509127},
year = {2025}
}

@article{reviewPaltiel,
  author    = {Naaman, Ron and Paltiel, Yossi and Waldeck, David H.},
  title     = {Chiral Molecules and the Spin Selectivity Effect},
  journal   = {The Journal of Physical Chemistry Letters},
  volume    = {11},
  number    = {9},
  pages     = {3660--3666},
  year      = {2020},
  month     = {05},
  day       = {07},
  doi       = {10.1021/acs.jpclett.0c00474},
  url       = {https://doi.org/10.1021/acs.jpclett.0c00474},
  publisher = {American Chemical Society}
}

@article{aragones2022magnetoresistive,
author = {Aragonès, Albert C. and Aravena, Daniel and Ugalde, Jesús M. and Medina, Ernesto and Gutierrez, Rafael and Ruiz, Eliseo and Mujica, Vladimiro and Díez-Pérez, Ismael},
title = {Magnetoresistive Single-Molecule Junctions: the Role of the Spinterface and the CISS Effect},
journal = {Israel Journal of Chemistry},
volume = {62},
number = {11-12},
pages = {e202200090},
doi = {https://doi.org/10.1002/ijch.202200090},
url = {https://onlinelibrary.wiley.com/doi/abs/10.1002/ijch.202200090},
year = {2022}
}

@article{kettner2018chirality,
  author    = {Kettner, Matthias and Maslyuk, Volodymyr V. and N\"{u}renberg, Daniel and Seibel, Johannes and Gutierrez, Rafael and Cuniberti, Gianaurelio and Ernst, Karl-Heinz and Zacharias, Helmut},
  title     = {Chirality-Dependent Electron Spin Filtering by Molecular Monolayers of Helicenes},
  journal   = {The Journal of Physical Chemistry Letters},
  volume    = {9},
  number    = {8},
  pages     = {2025--2030},
  year      = {2018},
  month     = {04},
  day       = {19},
  doi       = {10.1021/acs.jpclett.8b00208},
  url       = {https://doi.org/10.1021/acs.jpclett.8b00208},
  publisher = {American Chemical Society}
}

@article{calavalle2022gate,
  author    = {Calavalle, Francesco and Su\'{a}rez-Rodr\'{i}guez, Manuel and Mart\'{i}n-Garc\'{i}a, Beatriz and Johansson, Annika and Vaz, Diogo C. and Yang, Haozhe and Maznichenko, Igor V. and Ostanin, Sergey and Mateo-Alonso, Aurelio and Chuvilin, Andrey and Mertig, Ingrid and Gobbi, Marco and Casanova, F\`{e}lix and Hueso, Luis E.},
  title     = {Gate-tuneable and chirality-dependent charge-to-spin conversion in tellurium nanowires},
  journal   = {Nature Materials},
  volume    = {21},
  number    = {5},
  pages     = {526--532},
  year      = {2022},
  month     = {05},
  day       = {01},
  issn      = {1476-4660},
  doi       = {10.1038/s41563-022-01211-7},
  url       = {https://doi.org/10.1038/s41563-022-01211-7}
}

@article{aiello2022chirality,
  author    = {Aiello, Clarice D. and Abendroth, John M. and Abbas, Muneer and Afanasev, Andrei and Agarwal, Shivang and Banerjee, Amartya S. and Beratan, David N. and Belling, Jason N. and Berche, Bertrand and Botana, Antia and Caram, Justin R. and Celardo, Giuseppe Luca and Cuniberti, Gianaurelio and Garcia-Etxarri, Aitzol and Dianat, Arezoo and Diez-Perez, Ismael and Guo, Yuqi and Gutierrez, Rafael and Herrmann, Carmen and Hihath, Joshua and Kale, Suneet and Kurian, Philip and Lai, Ying-Cheng and Liu, Tianhan and Lopez, Alexander and Medina, Ernesto and Mujica, Vladimiro and Naaman, Ron and Noormandipour, Mohammadreza Caps and Palma, Julio L. and Paltiel, Yossi and Petuskey, William and Ribeiro-Silva, Jo\~{a}o Carlos and S\'{a}enz, Juan Jos\'{e} and Santos, Elton J. G. and Solyanik-Gorgone, Maria and Sorger, Volker J. and Stemer, Dominik M. and Ugalde, Jesus M. and Valdes-Curiel, Ana and Varela, Solmar and Waldeck, David H. and Wasielewski, Michael R. and Weiss, Paul S. and Zacharias, Helmut and Wang, Qing Hua},
  title     = {A Chirality-Based Quantum Leap},
  journal   = {ACS Nano},
  volume    = {16},
  number    = {4},
  pages     = {4989--5035},
  year      = {2022},
  month     = {04},
  day       = {26},
  issn      = {1936-0851},
  doi       = {10.1021/acsnano.1c01347},
  url       = {https://doi.org/10.1021/acsnano.1c01347},
  publisher = {American Chemical Society}
}

@article{inui2020chirality,
  title = {Chirality-Induced Spin-Polarized State of a Chiral Crystal ${\mathrm{CrNb}}_{3}{\mathrm{S}}_{6}$},
  author = {Inui, Akito and Aoki, Ryuya and Nishiue, Yuki and Shiota, Kohei and Kousaka, Yusuke and Shishido, Hiroaki and Hirobe, Daichi and Suda, Masayuki and Ohe, Jun-ichiro and Kishine, Jun-ichiro and Yamamoto, Hiroshi M. and Togawa, Yoshihiko},
  journal = {Phys. Rev. Lett.},
  volume = {124},
  issue = {16},
  pages = {166602},
  numpages = {6},
  year = {2020},
  month = {Apr},
  publisher = {American Physical Society},
  doi = {10.1103/PhysRevLett.124.166602},
  url = {https://link.aps.org/doi/10.1103/PhysRevLett.124.166602}
}

@article{guo2012spin,
  title = {Spin-Selective Transport of Electrons in DNA Double Helix},
  author = {Guo, Ai-Min and Sun, Qing-feng},
  journal = {Phys. Rev. Lett.},
  volume = {108},
  issue = {21},
  pages = {218102},
  numpages = {5},
  year = {2012},
  month = {May},
  publisher = {American Physical Society},
  doi = {10.1103/PhysRevLett.108.218102},
  url = {https://link.aps.org/doi/10.1103/PhysRevLett.108.218102}
}

@article{hawke2010electronic,
  author    = {Hawke, L. G. D. and Kalosakas, G. and Simserides, C.},
  title     = {Electronic parameters for charge transfer along {DNA}},
  journal   = {The European Physical Journal E},
  volume    = {32},
  number    = {3},
  pages     = {291--305},
  year      = {2010},
  month     = {07},
  day       = {01},
  issn      = {1292-895X},
  doi       = {10.1140/epje/i2010-10650-y},
  url       = {https://doi.org/10.1140/epje/i2010-10650-y}
}

@article{yang2020detecting,
  author    = {Yang, Xu and van der Wal, Caspar H. and van Wees, Bart J.},
  title     = {Detecting Chirality in Two-Terminal Electronic Nanodevices},
  journal   = {Nano Letters},
  volume    = {20},
  number    = {8},
  pages     = {6148--6154},
  year      = {2020},
  month     = {08},
  day       = {12},
  issn      = {1530-6984},
  doi       = {10.1021/acs.nanolett.0c02417},
  url       = {https://doi.org/10.1021/acs.nanolett.0c02417},
  publisher = {American Chemical Society}
}

@article{varela2016effective,
  title = {Effective spin-orbit couplings in an analytical tight-binding model of DNA: Spin filtering and chiral spin transport},
  author = {Varela, Solmar and Mujica, Vladimiro and Medina, Ernesto},
  journal = {Phys. Rev. B},
  volume = {93},
  issue = {15},
  pages = {155436},
  numpages = {16},
  year = {2016},
  month = {Apr},
  publisher = {American Physical Society},
  doi = {10.1103/PhysRevB.93.155436},
  url = {https://link.aps.org/doi/10.1103/PhysRevB.93.155436}
}

@article{varela2024optical,
    author = {Varela, Solmar and Gutierrez, Rafael and Cuniberti, Gianaurelio and Medina, Ernesto and Mujica, Vladimiro},
    title = {Chiral spin selectivity and chiroptical activity in helical molecules},
    journal = {The Journal of Chemical Physics},
    volume = {161},
    number = {11},
    pages = {114111},
    year = {2024},
    month = {09},
    issn = {0021-9606},
    doi = {10.1063/5.0227365},
    url = {https://doi.org/10.1063/5.0227365}
}

@article{bloom2024chiral,
  author    = {Bloom, Brian P. and Paltiel, Yossi and Naaman, Ron and Waldeck, David H.},
  title     = {Chiral Induced Spin Selectivity},
  journal   = {Chemical Reviews},
  volume    = {124},
  number    = {4},
  pages     = {1950--1991},
  year      = {2024},
  month     = {02},
  day       = {28},
  issn      = {0009-2665},
  doi       = {10.1021/acs.chemrev.3c00661},
  url       = {https://doi.org/10.1021/acs.chemrev.3c00661},
  publisher = {American Chemical Society}
}

@article{ray1999asymmetric,
author = {K. Ray  and S. P. Ananthavel  and D. H. Waldeck  and R. Naaman },
title = {Asymmetric Scattering of Polarized Electrons by Organized Organic Films of Chiral Molecules},
journal = {Science},
volume = {283},
number = {5403},
pages = {814-816},
year = {1999},
doi = {10.1126/science.283.5403.814},
URL = {https://www.science.org/doi/abs/10.1126/science.283.5403.814}}

@article{vzutic2004spintronics,
  title = {Spintronics: Fundamentals and applications},
  author = {\ifmmode \check{Z}\else \v{Z}\fi{}uti\ifmmode \acute{c}\else \'{c}\fi{}, Igor and Fabian, Jaroslav and Das Sarma, S.},
  journal = {Rev. Mod. Phys.},
  volume = {76},
  issue = {2},
  pages = {323--410},
  numpages = {0},
  year = {2004},
  month = {Apr},
  publisher = {American Physical Society},
  doi = {10.1103/RevModPhys.76.323},
  url = {https://link.aps.org/doi/10.1103/RevModPhys.76.323}
}

@article{gohler2011spin,
author = {B. Göhler  and V. Hamelbeck  and T. Z. Markus  and M. Kettner  and G. F. Hanne  and Z. Vager  and R. Naaman  and H. Zacharias },
title = {Spin Selectivity in Electron Transmission Through Self-Assembled Monolayers of Double-Stranded DNA},
journal = {Science},
volume = {331},
number = {6019},
pages = {894-897},
year = {2011},
doi = {10.1126/science.1199339},
URL = {https://www.science.org/doi/abs/10.1126/science.1199339}
}

@article{geyer2019chirality,
  author    = {Geyer, Matthias and Gutierrez, Rafael and Mujica, Vladimiro and Cuniberti, Gianaurelio},
  title     = {Chirality-Induced Spin Selectivity in a Coarse-Grained Tight-Binding Model for Helicene},
  journal   = {The Journal of Physical Chemistry C},
  volume    = {123},
  number    = {44},
  pages     = {27230--27241},
  year      = {2019},
  month     = {11},
  day       = {07},
  issn      = {1932-7447},
  doi       = {10.1021/acs.jpcc.9b07764},
  url       = {https://doi.org/10.1021/acs.jpcc.9b07764},
  publisher = {American Chemical Society}
}

@article{diaz2018thermal,
  author    = {D\'{i}az, Elena and Dom\'{i}nguez-Adame, Francisco and Gutierrez, Rafael and Cuniberti, Gianaurelio and Mujica, Vladimiro},
  title     = {Thermal Decoherence and Disorder Effects on Chiral-Induced Spin Selectivity},
  journal   = {The Journal of Physical Chemistry Letters},
  volume    = {9},
  number    = {19},
  pages     = {5753--5758},
  year      = {2018},
  month     = {10},
  day       = {04},
  doi       = {10.1021/acs.jpclett.8b02196},
  url       = {https://doi.org/10.1021/acs.jpclett.8b02196},
  publisher = {American Chemical Society}
}

@article{yeganeh2009chiral,
    author = {Yeganeh, Sina and Ratner, Mark A. and Medina, Ernesto and Mujica, Vladimiro},
    title = {Chiral electron transport: Scattering through helical potentials},
    journal = {The Journal of Chemical Physics},
    volume = {131},
    number = {1},
    pages = {014707},
    year = {2009},
    month = {07},
    issn = {0021-9606},
    doi = {10.1063/1.3167404},
    url = {https://doi.org/10.1063/1.3167404},
}

@article{mena2024minimal,
doi = {10.1088/1742-5468/ad613b},
url = {https://doi.org/10.1088/1742-5468/ad613b},
year = {2024},
month = {aug},
publisher = {IOP Publishing},
volume = {2024},
number = {8},
pages = {084001},
author = {Mena, Miguel and Varela, Solmar and Berche, Bertrand and Medina, Ernesto},
title = {Minimal model for chirally induced spin selectivity: spin-orbit coupling, tunneling and decoherence},
journal = {Journal of Statistical Mechanics: Theory and Experiment}
}

@article{ShiNiu2006,
  title = {Proper Definition of Spin Current in Spin-Orbit Coupled Systems},
  author = {Shi, Junren and Zhang, Ping and Xiao, Di and Niu, Qian},
  journal = {Phys. Rev. Lett.},
  volume = {96},
  issue = {7},
  pages = {076604},
  numpages = {4},
  year = {2006},
  month = {Feb},
  publisher = {American Physical Society},
  doi = {10.1103/PhysRevLett.96.076604},
  url = {https://link.aps.org/doi/10.1103/PhysRevLett.96.076604}
}

@article{farago1980spin,
doi = {10.1088/0022-3700/13/18/004},
url = {https://doi.org/10.1088/0022-3700/13/18/004},
year = {1980},
month = {sep},
publisher = {},
volume = {13},
number = {18},
pages = {L567},
author = {P S Farago},
title = {Spin-dependent features of electron scattering from optically active molecules},
journal = {Journal of Physics B: Atomic and Molecular Physics},
}

@article{mayer1995experimental,
  title = {Experimental Verification of Electron Optic Dichroism},
  author = {Mayer, Stefan and Kessler, Joachim},
  journal = {Phys. Rev. Lett.},
  volume = {74},
  issue = {24},
  pages = {4803--4806},
  numpages = {0},
  year = {1995},
  month = {Jun},
  publisher = {American Physical Society},
  doi = {10.1103/PhysRevLett.74.4803},
  url = {https://link.aps.org/doi/10.1103/PhysRevLett.74.4803}
}

@phdthesis{PMendieta,
  author = {Mendieta, Pablo},
  title = {Spin Transport Hamiltonians on Molecular Helices: Derivations from Symmetry},
  school = {Universidad San Francisco de Quito},
  year = {2024},
  type = {Undergraduate Honors Thesis},
  url = {https://repositorio.usfq.edu.ec/jspui/bitstream/23000/14170/1/206877.pdf}
}

@article{kochan2017model,
  title = {Model spin-orbit coupling Hamiltonians for graphene systems},
  author = {Kochan, Denis and Irmer, Susanne and Fabian, Jaroslav},
  journal = {Phys. Rev. B},
  volume = {95},
  issue = {16},
  pages = {165415},
  numpages = {19},
  year = {2017},
  month = {Apr},
  publisher = {American Physical Society},
  doi = {10.1103/PhysRevB.95.165415},
  url = {https://link.aps.org/doi/10.1103/PhysRevB.95.165415}
}

@article{michaeli2019origin,
  author    = {Michaeli, Karen and Naaman, Ron},
  title     = {Origin of Spin-Dependent Tunneling Through Chiral Molecules},
  journal   = {The Journal of Physical Chemistry C},
  volume    = {123},
  number    = {27},
  pages     = {17043--17048},
  year      = {2019},
  month     = {07},
  day       = {11},
  issn      = {1932-7447},
  doi       = {10.1021/acs.jpcc.9b05020},
  url       = {https://doi.org/10.1021/acs.jpcc.9b05020},
  publisher = {American Chemical Society}
}

@article{matityahu2016spin,
  title = {Spin-dependent transport through a chiral molecule in the presence of spin-orbit interaction and nonunitary effects},
  author = {Matityahu, Shlomi and Utsumi, Yasuhiro and Aharony, Amnon and Entin-Wohlman, Ora and Balseiro, Carlos A.},
  journal = {Phys. Rev. B},
  volume = {93},
  issue = {7},
  pages = {075407},
  numpages = {10},
  year = {2016},
  month = {Feb},
  publisher = {American Physical Society},
  doi = {10.1103/PhysRevB.93.075407},
  url = {https://link.aps.org/doi/10.1103/PhysRevB.93.075407}
}

@article{liu2020linear,
  author    = {Liu, Tianhan and Wang, Xiaolei and Wang, Hailong and Shi, Gang and Gao, Fan and Feng, Honglei and Deng, Haoyun and Hu, Longqian and Lochner, Eric and Schlottmann, Pedro and von Moln\'{a}r, Stephan and Li, Yongqing and Zhao, Jianhua and Xiong, Peng},
  title     = {Linear and Nonlinear Two-Terminal Spin-Valve Effect from Chirality-Induced Spin Selectivity},
  journal   = {ACS Nano},
  volume    = {14},
  number    = {11},
  pages     = {15983--15991},
  year      = {2020},
  month     = {11},
  day       = {24},
  issn      = {1936-0851},
  doi       = {10.1021/acsnano.0c07438},
  url       = {https://doi.org/10.1021/acsnano.0c07438},
  publisher = {American Chemical Society}
}

@article{nitzan2001electron,
  author    = {Nitzan, Abraham},
  title     = {Electron Transmission Through Molecules and Molecular Interfaces},
  journal   = {Annual Review of Physical Chemistry},
  volume    = {52},
  number    = {1},
  pages     = {681--750},
  year      = {2001},
  month     = {10},
  issn      = {0066-426X},
  doi       = {10.1146/annurev.physchem.52.1.681},
  url       = {https://doi.org/10.1146/annurev.physchem.52.1.681},
  publisher = {Annual Reviews}
}

@article{Cattena2010,
  title = {Crucial role of decoherence for electronic transport in molecular wires: Polyaniline as a case study},
  author = {Cattena, Carlos J. and Bustos-Mar\'un, Ra\'ul A. and Pastawski, Horacio M.},
  journal = {Phys. Rev. B},
  volume = {82},
  issue = {14},
  pages = {144201},
  numpages = {10},
  year = {2010},
  month = {Oct},
  publisher = {American Physical Society},
  doi = {10.1103/PhysRevB.82.144201},
  url = {https://link.aps.org/doi/10.1103/PhysRevB.82.144201}
}

@article{NaamanRacemicSeparation,
    author = {Naaman, Ron and Waldeck, David H. and Paltiel, Yossi},
    title = {Chiral molecules-ferromagnetic interfaces, an approach towards spin
          controlled interactions},
    journal = {Applied Physics Letters},
    volume = {115},
    number = {13},
    pages = {133701},
    year = {2019},
    month = {09},
    doi = {10.1063/1.5125034},
    url = {https://doi.org/10.1063/1.5125034},
}

@article{naaman2022chiral,
  author    = {Naaman, Ron and Paltiel, Yossi and Waldeck, David H.},
  title     = {Chiral Induced Spin Selectivity and Its Implications for Biological Functions},
  journal   = {Annual Review of Biophysics},
  volume    = {51},
  number    = {1},
  pages     = {99--114},
  year      = {2022},
  month     = {05},
  day       = {09},
  issn      = {1936-122X},
  doi       = {10.1146/annurev-biophys-083021-070400},
  url       = {https://doi.org/10.1146/annurev-biophys-083021-070400},
  publisher = {Annual Reviews}
}

@article{naaman2019chiral,
  author    = {Naaman, Ron and Paltiel, Yossi and Waldeck, David H.},
  title     = {Chiral molecules and the electron spin},
  journal   = {Nature Reviews Chemistry},
  volume    = {3},
  number    = {4},
  pages     = {250--260},
  year      = {2019},
  month     = {04},
  doi       = {10.1038/s41570-019-0087-1},
  url       = {https://doi.org/10.1038/s41570-019-0087-1},
  publisher = {Nature Publishing Group}
}

@book{harrison1989electronic,
author = {Harrison, Walter A.},
address = {New York},
edition = {Dover ed.},
isbn = {9781621986423},
publisher = {Dover Publications},
title = {Electronic structure and the properties of solids : the physics of the chemical bond},
url = {https://link.ezproxy.neu.edu/login?url=https://app.knovel.com/hotlink/toc/id:kpESPSTPC1/electronic-structure-and?kpromoter=marc},
year = {1989 - 1980},
}

@article{fransson2020vibrational,
  title = {Vibrational origin of exchange splitting and ''chiral-induced spin selectivity},
  author = {Fransson, J.},
  journal = {Phys. Rev. B},
  volume = {102},
  issue = {23},
  pages = {235416},
  numpages = {8},
  year = {2020},
  month = {Dec},
  publisher = {American Physical Society},
  doi = {10.1103/PhysRevB.102.235416},
  url = {https://link.aps.org/doi/10.1103/PhysRevB.102.235416}
}

@Article{Gupta2026CISSReview,
author ="Kumar, Anil and Gupta, Anu",
title  ="Chirality-induced spin selectivity: an interdisciplinary perspective from chemical physics to biology",
journal  ="Phys. Chem. Chem. Phys.",
year  ="2026",
volume  ="28",
issue  ="6",
pages  ="3812-3826",
publisher  ="The Royal Society of Chemistry",
doi  ="10.1039/D5CP04185F",
url  ="http://dx.doi.org/10.1039/D5CP04185F"
}

@article{Miwa2024VibrationalCISS,
author = {Shinji Miwa  and Tatsuya Yamamoto  and Takashi Nagata  and Shoya Sakamoto  and Kenta Kimura  and Masanobu Shiga  and Weiguang Gao  and Hiroshi M. Yamamoto  and Keiichi Inoue  and Taishi Takenobu  and Takayuki Nozaki  and Tatsuhiko Ohto },
title = {Spin polarization driven by molecular vibrations leads to enantioselectivity in chiral molecules},
journal = {Science Advances},
volume = {11},
number = {44},
pages = {eadv5220},
year = {2025},
doi = {10.1126/sciadv.adv5220},
URL = {https://www.science.org/doi/abs/10.1126/sciadv.adv5220}}

@article{Yoda2015HelicalMagnetization,
  author = {Yoda, T. and Yokoyama, T. and Murakami, S.},
  title = {Current-induced orbital and spin magnetizations in crystals with helical structure},
  journal = {Sci. Rep.},
  year = {2015},
  volume = {5},
  pages = {12024},
  doi = {10.1038/srep12024}
}

@article{Yoda2018OrbitalEdelstein,
  author    = {Yoda, Taiki and Yokoyama, Takehito and Murakami, Shuichi},
  title     = {Orbital Edelstein Effect as a Condensed-Matter Analog of Solenoids},
  journal   = {Nano Letters},
  volume    = {18},
  number    = {2},
  pages     = {916--920},
  year      = {2018},
  month     = {02},
  day       = {14},
  issn      = {1530-6984},
  doi       = {10.1021/acs.nanolett.7b04300},
  url       = {https://doi.org/10.1021/acs.nanolett.7b04300},
  publisher = {American Chemical Society}
}

@article{Cysne2025Orbitronics,
  author    = {Cysne, Tarik P. and Canonico, Luis M. and Costa, Marcio and Muniz, R. B. and Rappoport, Tatiana G.},
  title     = {Orbitronics in two-dimensional materials},
  journal   = {npj Spintronics},
  volume    = {3},
  number    = {1},
  pages     = {39},
  year      = {2025},
  month     = {10},
  day       = {01},
  issn      = {2948-2119},
  doi       = {10.1038/s44306-025-00103-1},
  url       = {https://doi.org/10.1038/s44306-025-00103-1}
}

@article{Wang2024OrbitronicsReview,
author = {Wang, Ping and Chen, Feng and Yang, Yuhe and Hu, Shuai and Li, Yue and Wang, Wenhong and Zhang, Delin and Jiang, Yong},
title = {Orbitronics: Mechanisms, Materials and Devices},
journal = {Advanced Electronic Materials},
volume = {11},
number = {5},
pages = {2400554},
doi = {https://doi.org/10.1002/aelm.202400554},
url = {https://advanced.onlinelibrary.wiley.com/doi/abs/10.1002/aelm.202400554},
year = {2025}
}

@article{Sala2022OrbitalConversion,
  title = {Giant orbital Hall effect and orbital-to-spin conversion in $3d$, $5d$, and $4f$ metallic heterostructures},
  author = {Sala, Giacomo and Gambardella, Pietro},
  journal = {Phys. Rev. Res.},
  volume = {4},
  issue = {3},
  pages = {033037},
  numpages = {14},
  year = {2022},
  month = {Jul},
  publisher = {American Physical Society},
  doi = {10.1103/PhysRevResearch.4.033037},
  url = {https://link.aps.org/doi/10.1103/PhysRevResearch.4.033037}
}

@article{Busch2023OrbitalHall,
  title = {Orbital Hall effect and orbital edge states caused by $s$ electrons},
  author = {Busch, Oliver and Mertig, Ingrid and G\"obel, B\"orge},
  journal = {Phys. Rev. Res.},
  volume = {5},
  issue = {4},
  pages = {043052},
  numpages = {9},
  year = {2023},
  month = {Oct},
  publisher = {American Physical Society},
  doi = {10.1103/PhysRevResearch.5.043052},
  url = {https://link.aps.org/doi/10.1103/PhysRevResearch.5.043052}
}

\end{document}